\documentclass[preprint, superscriptaddress, showpacs,preprintnumbers,amsmath,amssymb,prb,endfloats*]{revtex4}
\usepackage{graphicx}

\begin{document}
\pdfoutput=1
\thispagestyle{empty}

\title{Gradient of the Casimir force between Au surfaces of a sphere
and a plate measured using atomic force microscope in a frequency
shift technique}

\author{C.-C.~Chang}
\affiliation{Department of Physics and
Astronomy, University of California, Riverside, California 92521,
USA}

\author{ A.~A.~Banishev}
\affiliation{Department of Physics and
Astronomy, University of California, Riverside, California 92521,
USA}

\author{R.~Castillo-Garza}
\affiliation{Department of Physics and
Astronomy, University of California, Riverside, California 92521,
USA}

\author{G.~L.~Klimchitskaya}
\affiliation{North-West Open Technical University,
Polustrovsky Avenue 59,
St.Petersburg, 195597, Russia}

\author{V.~M.~Mostepanenko}
\affiliation{Central Astronomical Observatory
of the Russian Academy of Sciences,
Pulkovskoye chaussee 65/1,
St.Petersburg, 196140, Russia }

\author{U.~Mohideen}
\affiliation{Department of Physics and
Astronomy, University of California, Riverside, California 92521,
USA}

\begin{abstract}
We present measurement results for the gradient of the
Casimir force between an Au-coated sphere and an Au-coated
plate obtained by means of an atomic force microscope operated
in a frequency shift technique. This experiment was performed
at a pressure of $3\times 10^{-8}\,$Torr with hollow glass
sphere of $41.3\,\mu$m radius. Special attention is paid to
electrostatic calibrations including the problem of electrostatic
patches. All calibration parameters are shown to be separation-independent
after the corrections for mechanical drift are included. The gradient of
the Casimir force was measured in two ways with applied compensating
voltage to the plate and with different applied voltages and subsequent
subtraction of electric forces. The obtained mean gradients are shown
to be in mutual agreement and in agreement with previous experiments
performed using a micromachined oscillator.
The obtained data are compared with theoretical predictions of the Lifshitz
theory including corrections beyond the proximity force approximation.
An independent comparison with no fitting parameters demonstrated that
the Drude model approach is excluded by the data at a 67\% confidence level
over the separation region from 235 to 420\,nm. The theoretical approach
using the generalized plasma-like model is shown to be consistent with
the data over the entire measurement range.
Corrections due to the nonlinearity of oscillator are calculated and
the application region of the linear regime is determined.
A conclusion is made
that the results of several performed experiments call for a thorough
analysis of the basics of the theory of dispersion forces.

\end{abstract}

\maketitle

\section{Introduction}

Modern measurements of the Casimir force\cite{1} have been
actively pursued since
1997 (reviews\cite{2,3} contain the description of all
experiments performed up to 2001 and 2009, respectively).
In this period it has been conclusively demonstrated that the Casimir
force can be reproducibly measured using modern laboratory
techniques.
The obtained results have found prospective
applications ranging from nanotechnology\cite{4,5,6} to constraining
parameters of fundamental physical theories beyond the standard
model.\cite{c1,c2,8b,7,8,8a}

The Lifshitz theory\cite{9,10} of the
van der Waals and Casimir forces has been applied to two semispaces
made of real materials at
nonzero temperature\cite{11,12,13} and generalized for interacting
surfaces of arbitrary shape.\cite{14,15,16}
The comparison between the measurement data and computations using the
Lifshitz theory with particular dielectric permittivities has
shown areas of disagreement.
It was found that
theoretical predictions obtained using the dielectric permittivity
of the Drude model are excluded by the data of three precise
experiments\cite{17,18,19,20} on indirect dynamic determination of the Casimir
pressure between two parallel metallic plates
performed in the configuration of a sphere above a plate
by means of a  micromachined
oscillator at separations from 0.16 to $0.75\,\mu$m
(note that in the modern phase of measurements of the Casimir
force the experimental configuration of two parallel plates
was used only in Ref.\cite{20a}).
The same data were found
consistent with the Lifshitz theory combined with the generalized
plasma-like model.\cite{20} Coincidently it was proven\cite{3,21} that
the Lifshitz theory combined with the plasma model satisfies the third
law of thermodynamics (the Nernst heat theorem), but violates this
fundamental physical principle when the dielectric permittivity of the
Drude model for metals with perfect crystal lattices is used.
Keeping in mind that the dielectric response of metals for real
electromagnetic fields is correctly described by the Drude model,
whereas the plasma model is an approximation applicable only at
sufficiently high frequencies, these results initiated continuing
discussions.\cite{2,22,23,24}
Specifically, it was suggested to carefully check all approximations
used in the computations, in particular, the proximity force approximation
(PFA), reconsider the role of such background effects as
electrostatic patch potentials
and surface roughness, and to determine the role of variations in
optical data of the Au films.

Results similar in spirit were obtained with dielectric surfaces.
Thus, measurements of the difference in the Casimir forces between
an Au-coated sphere and Si plate in the presence and in the absence
of a laser pulse  on a plate
measured by means of an atomic force microscope (AFM) at short separations
were found to be consistent with the Lifshitz theory
when neglecting the
dc conductivity of dielectric Si in the dark phase.\cite{25,26}
The same measurement data exclude the Lifshitz theory with the
dc conductivity of Si  taken
into account in the absence of a laser
pulse.\cite{25,26}
Measurements of the thermal Casimir-Polder force between ${}^{87}$Rb
atoms belonging to the Bose-Einstein condensate and SiO${}_2$
plate\cite{27} agree well with the Lifshitz theory neglecting the
dc conductivity of SiO${}_2$, but were found\cite{28} in clear
disagreement with the same theory including the dc conductivity in
the computations.
On the theoretical side, it was demonstrated\cite{29} that the Lifshitz
theory for dielectrics with included dc conductivity violates the
Nernst heat theorem. Theoretical and experimental advances in the
Casimir effect, including the unresolved problems, are discussed
in the book,\cite{30} whereas reviews\cite{31,32} present details
of  experiments with fluids and semiconductors, respectively.

Recently two more experiments were performed. The first of them\cite{33}
 was performed by means of a torsion pendulum.
It reports observation of the thermal Casimir force at large
separations $0.7-7.3\,\mu$m between a spherical
lens of $R=15.6\,$cm radius of curvature and a plane plate both coated
with Au, as predicted by the Lifshitz theory using the Drude
model. This experiment did not directly measure the Casimir force but up to
an order of magnitude larger total force presumably determined by large
electrostatic patches.\cite{33} The Casimir force was extracted using a fitting
procedure with two fitting parameters.  The results of this experiment
are in contradiction with two earlier torsion-pendulum
experiments performed at large separations,\cite{36,37}
and with dynamic measurements by means of micromachined
oscillator.\cite{17,18,19,20}
These results have been called into question in the
literature.\cite{35,35a}
The second of two recent experiments measured the Casimir force
between an Au-coated sphere and an indium tin oxide (ITO) plate using
an AFM in the static mode.\cite{34,34a} It was shown that the
 UV treatment
of an ITO plate results in up to 35\% decrease in the magnitude of
the Casimir force. On the basis of the Lifshitz theory this was
explained by the Mott-Anderson phase
transition of the ITO plate under the influence of
UV treatment from metal to dielectric state, where the dc conductivity
of ITO was omitted in computations.
However, the inclusion of the dc conductivity
of the UV-treated sample results in the contradiction of theoretical
predictions with the experimental data.

Keeping in mind that the Lifshitz theory faces outstanding problems when
using the most natural and well tested models of dielectric response,
it is important to perform more experiments on measuring the
Casimir force between Au surfaces particularly using a laboratory setup
different in  technique and
preparation of the test bodies and from those
applied previously.
In this paper we report measurements of the gradient of the Casimir force
between an Au-coated hollow glass microsphere and an Au-coated sapphire
plate by means of the dynamic AFM operated in the frequency shift
technique (also referred to as frequency modulation).
Previously measurements of the Casimir interaction between Au surfaces,
allowing discrimination between predictions of the Drude and plasma models
at short separations, were performed by means of a micromachined
oscillator.\cite{17,18,19,20} The experiment\cite{38} with Au-coated
surfaces of a
sapphire disk and a polystyrene sphere using the static AFM was not enough
precise to discriminate between different theoretical predictions.
Similar AFM experiments have been performed in the dynamic mode using the phase
shift\cite{39,40a} and the amplitude shift\cite{40} techniques.
These have also not been precise enough to distinguish between the
various models.
In the frequency shift technique, the gradient of the Casimir force acting
on the cantilever modifies the resonant frequency and the corresponding
shift in frequency is recorded using a phase locked loop
(with application to AFM the frequency shift technique is
discussed in detail in Ref.\cite{40b}).
As a result, the frequency shift technique is free from limitations
inherent to the phase shift and amplitude shift techniques and leads to
a factor of 10 larger sensitivity. This allows discrimination between
different theoretical approaches to the thermal Casimir force using the AFM.

This experiment was performed at a lower pressure of $3\times 10^{-8}\,$Torr.
Much attention was paid to electrostatic calibration. Specifically, all
mechanical drifts were measured and the corresponding corrections in the
measured quantities were introduced. As a result, the residual potential
difference, the closest sphere-plate separation and the coefficient
converting the frequency shift into the force gradient were found independent
of separation distance. The numerical simulation of the electrostatic force
due to electrostatic
patch potentials was performed. It was shown that for both small and
large patches the residual potential between the sphere and the plate would
heavily depend on the separation for patch sizes of order or larger than this
separation. The absence of such dependence in our measurements confirms
the fact that the interacting regions of the test bodies used were free
from work function inhomogeneities of a size scale which may
significantly distort the
total force measured.

The gradient of the Casimir force between the sphere and the plate was
obtained in two different ways: from the total force (electric plus
Casimir) with different applied voltages and electric force subtracted
(44 measurement sets) and in an immediate way by the application of
only the compensating voltage to counterbalance the residual potential
difference (40 measurement sets). The mean gradients of the Casimir force
obtained in these two ways were found in good mutual agreement and in
agreement with the measurements by means of the micromachined
oscillator.\cite{19,20} The random, systematic and total experimental
errors at a 67\% confidence level have been determined.

The experimental data were compared with theoretical predictions of the
Lifshitz theory. In so doing the corrections to the PFA obtained from
exact calculations in the sphere-plate geometry were taken into account.
Computations of the gradient of the Casimir force were performed taking
the surface roughness into account. The roughness profiles were investigated
using an atomic force microscope. In computations the optical data for
the complex index of refraction of Au from different sources have been used.
The theoretical prediction using the Drude model approach (i.e., the
tabulated optical data for the imaginary part of dielectric permittivity
extrapolated to lower frequencies by the imaginary part of the Drude
model) is excluded by the results of our measurements at a 67\% confidence
level over the separation region from 235 to 420\,nm.
The Lifshitz theory combined with the generalized plasma-like model is shown
to be consistent with the measurement data over the entire measurement range
from 235 to 750\,nm. We have also studied the oscillator system used in the
experiment in the nonlinear regime and determined the application region of
the linear equation used to convert the frequency shift of the oscillator into
the gradient of the Casimir force.

The structure of the paper is as follows. In Sec.~II we describe the
experimental setup and scheme for dynamic measurements of the gradient
of the Casimir force in the frequency shift technique. Electrostatic
calibrations and related problems are considered in Sec.~III.
Here, we pay special attention to the problem of patch potentials by
performing numerical simulations of additional forces due to electrostatic
patches or contaminants. All calibrations are performed with account
of the mechanical drift in separation distances. Section~IV contains
measurement results for the gradient of the Casimir force.
We present the measurement data obtained in two ways, with applied
compensating voltage and with applied different voltages with subsequent
subtraction of the electric force.
In Sec.~V we perform the comparison between experiment and theory.
The role of nonlinear effects in dynamic measurements using an AFM is
discussed in Sec.~VI. In Sec.~VII the reader will find our conclusions
and discussion.

\section{Setup and scheme for dynamic measurements in the frequency shift
technique}

The main role in the setup for dynamic measurement of the gradient of the
Casimir force in the frequency shift technique [see Fig.~1(a)]
is played by the detection
system. This system consists of an AFM cantilever with attached sphere,
piezoelectric actuator, fiber interferometers, light source and phase
locked loop (PLL). The detection system was placed in a high vacuum
chamber [see Fig.~1(a)]. The high vacuum down to $10^{-9}\,$Torr was created and sustained
with the help of different pumps, valves, gauges  and various vacuum
feed-throughs. We begin the description of the setup with the vacuum system.

For the main vacuum chamber an $8^{\prime\prime}$ six-way stainless steel cross
was used. This chamber was mounted on a
$8^{\prime\prime}$ ion pump.
The chamber was first evacuated by a turbo-pump backed by an
oil-free dry scroll mechanical pump.  The first two pumps can achieve a vacuum down
to $2\times 10^{-7}\,$Torr. The ion pump allows to reach a vacuum of
$10^{-9}\,$Torr. The vacuum chamber was separated from the turbo and mechanical
pumps by a gate valve which can be closed when only the ion pump is to be
used. The low vacuum pressure less than $10^{-3}\,$Torr was measured with
a thermal-conductivity gauge. For measuring high vacuums of $10^{-9}\,$Torr
an ionization gauge was used. The main vacuum chamber was supported
on a damped optical table having a large mass to reduce the mechanical noise.
For the electrical connections to elements inside the vacuum chamber
D-type subminiature connections were used offering UHV feed-through with
25 pins which were hermetically sealed and insulated by means of glass
ceramic bonding. For optical connections a home-built optical fiber
feed-through was used made on a CF flange welded with a clean stainless
tube. A cladding-stripped 1550\,nm fiber was inserted through the steel
tube and sealed using Varian vacuum Torr seal.

We continue the description of the setup by
acquaintance with the fiber interferometers [see Fig.~1(b) for more details].
 Two fiber optic interferometers were used. One interferometer monitored
the cantilever oscillation. The second recorded the displacement of the
Au plate mounted on the AFM piezoelectric actuator.
For constructing the interferometer,
we used a 1550\,nm single mode fiber which has
extremely low bending loss and low splice loss compared to the standard
SMF-28 1550\,nm fiber. A super luminescent diode
with a wavelength of 1550\,nm, served as the light source for the
cantilever frequency measurement interferometer. The coherence length
of the diode was $66\,\mu$m. A short coherence length is necessary to avoid
noise from spurious interferences from unwanted reflections. An optical
isolator with FC-APC connectors joined the diode to
a 50/50 directional coupler.
We used the typical fused-tapered bi-conic coupler at 1550\,nm wavelength
with return-loss of --55dB relative to the input power. To reduce the
signal power attenuation, we avoided bulkhead connectors which usually
have $\sim0.3$dB power loss. A fiber coupled laser source
with a wavelength of 635\,nm was used
for the Au plate displacement interferometer.  In our experimental
setup, we used a commercial anodized black xyz-stage
[see Fig.~1(b)] to move the
fiber end vertically above and close to the cantilever. But the
anodizing  may significantly increase outgassing
rates because of its porous structure. Therefore special treatments
were done before placing the xyz-stage into the high vacuum chamber.
The xyz-stage was first disassembled and the parts scrubbed with strong
solution of detergent in an ultrasonic cleaner. They were then rinsed
with very hot water. Next they were immersed in a 10\% solution of sodium
hydroxide (NaOH) saturated with common salt (NaCl) at $80^{\circ}$C.
The parts
were then polished in a conventional wheel polishing machine. They were
then immersed in 10\% solution of hydrochloric acid to obtain a bright
finish. This was then followed by rinsing in DI water and reassembly
of the xyz stage using  powder-free disposable plastic gloves.
Finally the xyz stage was rinsed with acetone followed by ethyl
alcohol before insertion into the vacuum chamber.

The interferometer system [see Fig.~1(b)] used for measuring the
frequency shift consists of the following parts.
The output of the directional
couplers was measured with InGaAs photodetectors. A low noise photodetector
and amplifier system was fabricated for the cantilever interferometer.
To avoid any potential errors or noise from the divider or balancing
system, the InGaAs diode was coupled to an OPA627 low noise operational
amplifier (very high input impedance $\sim\!\!10^{13}\Omega$)
as a trans-impedance
amplifier. The output of the interference signal was fed into a
band-pass filter
 cascaded by low  and high
pass filter to cut off unwanted frequency bands. The high-pass filter
helped us to remove the noise in the excitation signal from
frequency modulation controller
and the low-pass filter removed $1/f$ and environmental noise. For frequency
demodulation we use a PLL (Nano Surf.).
The PLL frequency demodulator system combines a controller module and
detector module to measure the force gradient induced resonant frequency
shift. The output of the high pass filter was fed to the PLL for the frequency
demodulation. The output of the PLL was connected to a piezoelectric
transducer
 which drives the cantilever at its resonant frequency
$\omega=2\pi\times1.604\,$kHz
with a constant amplitude.
The oscillation amplitude of the cantilever was fixed at $<10\,$nm for all
separations. The output of the low pass filter was used to form a closed loop
proportional-integral-derivative  (PID)
controller and to maintain constant separation distance between the
fiber  end and the cantilever during the frequency modulation
detection. The distance was kept
constant using a piezoelectric actuator with the PID controller.

Now let us consider the detection system. As mentioned above,
the force gradient was measured between an Au-coated plate and
Au-coated sphere attached to the
cantilever [see Fig.~1(b)].
In difference with previous experiments, to achieve high
resonance frequencies (and therefore low noise) hollow glass microspheres
and stiffer rectangular Si cantilevers
 were used.  The Si cantilevers
are conductive, which is necessary for electrical contact to the sphere.
The surfaces of the hollow glass spheres are smooth as they are made from
liquid phase. We followed a special procedure for cleaning the spheres before
attachment to the cantilever. The sphere cleaning procedure removes organic
contaminants and debris from the surface. The first step in the cleaning
process was to make a 10\,ml solution of ethanol into which the spheres
were deposited. This solution was thoroughly mixed in a vortex mixer for
about 1 minute. Then, using a pipette, the alcohol was extracted from the
solution, leaving the spheres in the vial. To remove strongly attached
adsorbates and debris, a 10 ml solution of Hydrogen peroxide
(H${}_2$O${}_2$) was
added to the vial.  A vortex mixer was again used to mix for about 2\,min.
The O${}_2$ gas released from the solution removed any attached
debris from the
spheres. Next, the
H${}_2$O${}_2$ was extracted with a pipette, and the spheres were
immersed in 10\,ml of ethanol again. To completely separate the debris from
the spheres, we centrifuged the alcohol/sphere solution, for 10~minutes.
The large radius, clean hollow spheres float to the top of the mixture and
the debris sediment to the bottom.  We used a pipette to remove the spheres
at the top and place them in a pyrex petri dish to dry.  The dried spheres
were then mounted on the conductive Si cantilevers using Ag epoxy.
The cantilever-sphere system and  a 10\,mm in diameter sapphire disk
were coated with Au in an oil-free thermal evaporator
described elsewhere.\cite{41} The sphere cantilever system was
rotated to get a uniform coating. In contrast to previous experiments
only the tip of the cantilever with the sphere was coated with Au. Complete
 coating of the cantilever leads to large decrease in  its oscillation
Q-factor and loss of sensitivity. The thickness of
the Au coating was measured
to be $280\pm 1\,$nm using an AFM. The radius of the sphere was measured to
be $R=41.3\pm 0.2\,\mu$m using a scanning electron microscope.
Thus, the sphere coating was thicker and the sphere radius smaller than in
the static AFM experiment.\cite{38}

The cantilever with the sphere was placed in a specially fabricated holder
[see Fig.~1(b)]
containing two piezos (first connected to the PID loop and the second to
the PLL). The Au plate was mounted on top of a tube piezoelectric
actuator from a commercial
AFM capable of traveling a distance of $2.3\,\mu$m.
Ohmic contacts were made to the Au plate through a 1\,k$\Omega$ resistor.
To minimize electrical ground loops all the electrical ground connections
were unified to the AFM ground. The calibration of the tube
piezoelectric actuator was
done using the second interferometer and is described in previous
work.\cite{42}
To change the sphere plate distance using
the piezoelectric actuator
and avoid piezo drift and creep, a continuous 0.01\,Hz
triangular voltage signal was applied to the actuator. The chamber
was evacuated using the turbo pump [see Fig.~1(a)] to a pressure of
$2\times 10^{-7}\,$Torr. Next,
the ion pump [see Fig.~1(a)] was turned on and the turbo pump was isolated
from the chamber
by closing the $6^{\prime\prime}$ gate valve. When chamber reaches
$3\times 10^{-8}\,$Torr pressure,
 the experiment was started.

The measurement scheme allowing determination of the gradient of the
Casimir force is the following. In a dynamic experiment using the
frequency shift technique the total force gradient acting on the
cantilever due to interaction of the sphere and plate surfaces modifies
the natural resonance frequency of the oscillator. The corresponding change
in the frequency $\Delta\omega=\omega_r-\omega_0$, where $\omega_r$
is the resonance frequency in the presence of external force
$F_{\rm tot}(a)$, $a$ is the separation between the sphere and the plate,
is recorded by the PLL.
The total force $F_{\rm tot}(a)$ is the sum of the
electrostatic force $F_{\rm el}(a)$ and the Casimir force $F(a)$
\begin{equation}
F_{\rm tot}(a)=F_{\rm el}(a)+F(a).
\label{eq1}
\end{equation}

Note that even if no voltage is applied to the Au plate and the Au
sphere is grounded, there is some residual potential difference $V_0$
between the interacting bodies. This is caused by different connections
or work functions of sphere and plate materials from patches and
adsorbates on their surfaces. To perform the
electrostatic calibrations of our measurement system,
 the Au coated plate
was connected to a voltage supply
operating with $1\,\mu$V resolution.
Then 11 different voltages
$V_i$ in the range
from --87.4 to 32.6\,mV were applied to the Au plate, while the sphere
remained grounded.

Starting at the maximum separation, the plate was
moved towards the sphere and the corresponding frequency shift was
recorded at every 0.14\,nm.
As was mentioned above, the sphere-plate force acting on the
cantilever causes a change in the resonance frequency.
In the linear regime it is given by
\begin{equation}
\Delta\omega=-\frac{\omega_0}{2k}\,
\frac{\partial F_{\rm tot}(a)}{\partial a},
\label{eq2}
\end{equation}
\noindent
where $k$ is the spring
constant of the cantilever.
Equation (\ref{eq2}) is an approximate one. In Sec.~VI it is obtained
from the solution of a more complicated nonlinear problem and the
application region of this equation is determined.
The electric force entering Eq.~(\ref{eq1}) is expressed as
\begin{equation}
F_{\rm el}(a)=X(a,R)(V_i-V_0)^2,
\label{eq3}
\end{equation}
\noindent
where the function $X(a,R)$ can be written in the form\cite{30,43}
\begin{eqnarray}
&&
X(a,R)=2\pi\epsilon_0\sum_{n=1}^{\infty}
\frac{\coth\alpha-n\coth{n\alpha}}{\sinh{n\alpha}},
\nonumber \\
&&
 \cosh\alpha=1+\frac{a}{R}
\label{eq4}
\end{eqnarray}
\noindent
and $\epsilon_0$ is the permittivity of the free space.
For convenience in computations this function can be presented
approximately\cite{30,44} as the sum of powers $c_i(a/R)^i$
from $i=-1$ to $i=6$.
The absolute separation $a$ is measured between the zero levels of
the roughness on the surfaces of a sphere and a plate (see Sec.~V).
Experimentally the absolute separation is found as
$a=z_{\rm piezo}+z_0$, where
$z_{\rm piezo}$ is the plate movement due to the piezoelectric
actuator and $z_0$
is the point of closest approach between the Au sphere and Au plate
(note that in this experiment the separation at the closest approach
is much larger than the separation on contact of the two surfaces).

Substituting Eq.~(\ref{eq1}) in Eq.~(\ref{eq2}) and using Eq.~(\ref{eq3}),
one obtains
\begin{equation}
\Delta\omega=
-\beta (V_i-V_0)^2-C\frac{\partial F}{\partial a},
\label{eq5}
\end{equation}
\noindent
where $C\equiv C(k,\omega_0)=\omega_0/(2k)$ and
$\beta\equiv\beta(a,z_0,C,R)=C\partial X(a,R)/\partial a$.
Here, the first term on the right-hand side is associated with
the gradient of electrostatic
force caused by  a constant voltage applied to the plate
while the sphere remained grounded.
The second term is proportional to the gradient of the Casimir force.
Below Eq.~(\ref{eq5}) is used for the determination of
the gradients of the Casimir force from the measured data
for the frequency shifts.

\section{Electrostatic calibrations and related problems}

To determine the gradients of the Casimir force from Eq.~(\ref{eq5})
one needs to know the precise values of the involved parameters
$\beta$, $C$ and $z_0$. These can be determined by applying different
voltages $V_i$ to the plate and investigating the parabolic dependence
of the frequency shift expressed in arbitrary units on $V_i$.
The calibration procedure requires much care because errors
in calibration parameters due, for instance, mechanical drift result in
respective errors in the measured Casimir force or its gradient.
Electrostatic calibrations in measurements of the Casimir force have created
much discussion in the literature.\cite{45,46,47,48,49}
Specifically, in some cases it was observed\cite{45} that the residual
potential difference $V_0$ depends on separation: $V_0=V_0(a)$.
This was attributed to the probable role of the patch potentials
arising due to the polycrystal structure of metallic coatings\cite{51}
or dust and contaminants on the surfaces \cite{52} (for a pure Coulombian
interaction, similar to the applied voltages considered above, $V_0$ is
separation-independent). It was even speculated\cite{49} that patches
may render the experimental data\cite{19,20}
at distances below $1\,\mu$m compatible with theoretical
predictions based on the Drude model. Below we present the results of
numerical simulation of the electrostatic force due to patch potentials and
then describe the electrostatic calibrations.

\subsection{Numerical simulations of additional forces due to
electrostatic patches}

Here, we consider the force arising from a spatial distribution of
electrostatic potentials on the surface of an Au plate in close proximity
to an Au sphere for typical parameters of experimental interest.
A realistic variation of the patch potentials was chosen in the range
between $-90$ and 90\,mV.
 This was based on the maximum difference in the workfunctions between
the $\langle 100\rangle$ and the $\langle 111\rangle$ crystal
orientations of Au.\cite{53}
The electrostatic force between the Au plate with patches and
the grounded Au
sphere was numerically simulated using 3-dimensional
finite element analysis commercial software package
(COMSOL Multiphysics).

For this purpose an additional package of the AC/DC module of
COMSOL Multiphysics, which is software specifically designed for problems
with electrodynamics and electrostatics, was used. The physical system used
in the simulation was drawn with the CAD module of COMSOL Multiphysics
and the boundary conditions discussed below were assigned.

We first drew a $100\,\mu$m diameter sphere and
$120\times 120\times 20\,\mu\mbox{m}^3$ plate (see Fig.~2).
The sphere and plate were placed inside a rectangular box held at
constant potential. To represent the patches, square grids of various
sizes were made on the plate. The size of the grids varied between
300\,nm to 6000\,nm. Random potentials between --90 to +90\,mV were
assigned to the patches using a random number generator.  There was
no height difference between the patches and the surrounding.

After that, the governing equation, that needs to be solved, with the
given boundary conditions was defined. For the electrostatic
force between two metallic objects in vacuum with applied voltages
to the plate, the governing equation is the Poisson equation:
$\nabla\varepsilon_0\varepsilon_r\nabla{V}=0$,
where $\varepsilon_r$ is the relative permittivity set equal to unity and
$V$ is the potential. The numerical program solves the electric field
and potential at each coordinate in the 3-dimensional space.
It finds the force for a given configuration of potentials on the
surfaces. The force acting on the sphere was obtained by integrating
the Maxwell stress tensor along the sphere boundary.

The boundary conditions were as follows. The sphere, and the external
box were grounded (potentials were set equal to zero).
The sphere and plate were positioned at some distance $a$.
First the force only due to the patches, $F_{\rm pat}$, was calculated.
Here the potential applied to the plate is kept at zero and only the
patches have nonzero potentials. Next, the sphere-plate force for
various potentials $V$ applied to the plate, $F_V$, was calculated,
when the patch potentials were set to zero. The value of the applied
potential $V$ to the plate was varied till $F_V=F_{\rm pat}$.
This value of $V$ is equal to $V_0$, the potential that is necessary
to compensate the electrostatic force from the patches.
Zero electrostatic sphere-plate force was confirmed for the plate with
patches by applying a potential of $V_0$ to the plate.

To achieve the highest resolution in
solving the electrostatic problem we used the following
parameters for generation of the surface mesh:
1) The maximum element size is $10^{-7}$;
2) Maximum element size scaling factor is 1;
3) Element growth rate is 1.2;
4) Mesh curvature factor is 0.2;
5) Mesh curvature cutoff is 0.001;
6) Resolution of narrow regions is 1.
The latter parameter was used to control the number of elements
generated in a narrow region.
Among these parameters, 3)--6) are very important if a finer mesh is
required,  especially for complicated objects with sharp, narrow
edges or small holes.

Now we consider the results of the influence of the random distribution
of electrostatic patch potentials on the electrostatic force between
the gold sphere and plate. The investigations were performed for
periodic patches with random electrostatic potentials distribution
for different sphere-plate separations. The corresponding compensation
voltage $V_0$ when $F_V=F_{\rm pat}$  was found at every separation.

In the finite element analysis method the errors mainly result from
the discretization of the surface mesh elements. To obtain the necessary
precision and optimize the computation time, the number of mesh elements
was restricted to 243122. To calibrate the simulation, the theoretical
force was compared to that obtained from the numerical simulation.
In these checks a uniform plate with no patches was used.
The sphere and plate were placed at 100\,nm separation.
The voltage $V$ was applied to the plate and the corresponding $F_V$ was
found using the numerical simulation. The value was found to be precise
to within 0.23\% of the theoretical value. This was repeated for several
sphere-plate separation distances between 50--1000\,nm.

First the role of large patches was studied. Patch sizes of
$6\times 6\,\mu\mbox{m}^2$ were used on the plate. The distance between the
patches was kept at $6\,\mu$m. The residual potentials that will compensate
the electrostatic force of the patches were found for sphere-plate
separation distances from 50\,nm to 1000\,nm. These are shown
in Fig.~3(a) as solid squares. The residual potential was found to change
as a function of the separation distance. The change decreases with
separation and is very small at 1000\,nm even for these $6\,\mu$m patches.
Next the separation distance between the patches was
increased to $9\,\mu$m, while the
patch size remained fixed. The calculations were repeated as before
for different sphere-plate separation distances and results are shown
as solid circles in Fig.~3(a).
The same exercise was repeated for $12\,\mu$m distance between the
patches and the compensation voltage found is shown as solid triangles
in the same figure.
It is clear from the figure that while the values of the compensation
potential are different for different separation distances the
dependence of $V_0$ on separation is very similar. This shows that even
for these relatively large patches it is possible to obtain a region of
separation distance on the order of a micron where the compensation
potential is relatively independent of separation distance.

Next we investigated the role of smaller patches of size
$900\times 900\,\mbox{nm}^2$. The distances between the patches were fixed at
600\,nm. The simulation was repeated and the compensation voltage
was found for separation distances between 50\,nm and 1000\,nm.
The values are shown as solid triangles in Fig.~3(b).
The $V_0$ is found to vary as a function of the separation distance
from 9\,mV to 21\,mV. Then the patch size was decreased to
$600\times 600\,\mbox{nm}^2$ while the patch-patch distance remained fixed.
The simulations were repeated and the compensation voltage as a function
of separation is shown as solid circles in Fig.~3(b).
The smaller patches are seen to lead to smaller $V_0$ which also varies
less with separation distance. The variation seems to be correlated
with the patch size and the $V_0$ varies little for separation distances
greater than the patch size.

Then the patch size was further reduced to $300\times 300\,\mbox{nm}^2$ and
simulations repeated for the same patch-patch distance.
For these smaller patches the plate size was reduced to
$32\times 32\,\mu\mbox{m}^2$
to optimize the computation time. The results of the compensation voltages
are represented by solid squares in Fig.~3(b).
It is clear that the compensation voltage $V_0$ of the smaller patches
have a correspondingly weak dependence on the separation distance.
It is worth noting that the $300\times 300\,\mbox{nm}^2$ patches have a $V_0$
which is almost independent of the separation distance particularly
above 300\,nm. It should be mentioned that the crystallite sizes observed
on the Au coated plates used in experiments are even less than 300\,nm.
This implies that clean samples where the patch effects originate only
from crystallite size should show a $V_0$ that is independent of
sphere-plate separation distance. It should also be emphasized that
the values of $V_0$ between 8 and 9\,mV computed are similar to those
observed in experiments. This lends credence to the notion that the
patches present in experiments are of similar size and potential as
has been simulated.

At the same time, patches or surface contaminants of about $2\,\mu$m size,
like those considered recently\cite{49} in order to make the
experimental data\cite{19,20} compatible with theoretical predictions of the
Lifshitz theory combined with the Drude model, lead to a very strong
dependence of $V_0$ on the separation distance at separations up to
a few hundred nanometers [see the upper line marked by solid triangles
in Fig.~3(b)].
In several experiments\cite{17,18,19,20,33,34,34a,38,46}
it was experimentally found, however, that $V_0={\rm const}$ over
the entire measurement range.
This rules out the hypothesis\cite{49} that the deviation between the
experimental data\cite{19,20} and the Drude model predictions may be
attributed to the role of large patches. We will return to this point
in the next section in connection with our measurements.

\subsection{Calibration with account of mechanical drift}

Now we describe how the parameters of our measurements,
$V_0$, $z_0$ and $C$ were determined by means of electrostatic
calibrations. We begin with the determination of $V_0$.
{}From Eq.~(\ref{eq5}) it can be seen that the frequency shift
$\Delta\omega$ has a parabolic dependence on the voltage $V_i$
applied to the plate and reaches a maximum at $V_i=V_0$.
Thus, $V_0$ can be extracted from this dependence using
$\chi^2$-fitting procedure. The curvature of the parabolas is
related to $\beta$ and includes the closest separation and the
cantilever parameter $C$. Thus, these parameters can be also extracted
from the fit. In order to test for systematic errors in the fitting
parameters, the fitting procedure was repeated many times at
different distance ranges.

In the first step, 11 different voltages $V_i$ were applied to the plate and the
corresponding frequency shift signal due to the total force
gradient was measured.
 Then, we subtracted any drift of the frequency shift signal.
For this purpose we used the fact that for
separations larger than $2\,\mu$m the total force between the sphere
and plate is below the instrumental sensitivity. At these separations,
the noise is far greater than the signal and in the absence of systematic
errors the signal should average to zero. Then
the correction due to the drift
in sphere-plate separation can be measured. The effect of the drift can be
observed in Fig.~4(a), where 8 repeated measurements of the frequency
shift signal $S_{\Delta\omega}$
as a function of the sphere-plate separation change $\Delta z$ are
shown for same applied voltage to the plate. Drift causes the separation
 to increase by around 1\,nm in 700\,s, where 100\,s correspond
to the time taken to make the one measurement [note that positional precision
much better than 1\,nm is achieved in this experiment as observed
in Fig.~4(a)].
To calculate the drift rate, the change in position at one frequency shift
signal is plotted as a function of time as shown in Fig.~4(b).
This was repeated for 15 different signals and the average
drift rate was found to be 0.002 nm/s. The separation distances in all
measurements and between subsequent measurements
were corrected for this drift rate.

    After applying the drift correction the residual potential $V_0$
between the sphere and plate was found. The frequency shift signals at
every 1\,nm separation were found by interpolation (data
acquisition was done every 0.14\,nm).  From the parabolic dependence of the
electric force gradient shown in Eq.~(\ref{eq5}), $V_0$ can be
identified at the
position of the parabola maxima. The frequency shift signal was plotted
as a function of the applied voltages $V_i$ at every separation and the
corresponding $V_0$ and curvature of the parabola $\beta$ were found.
This $V_0$ is
shown in Fig.~5(a)  as a function of separation distance with a
step of 1\,nm. In Fig.~5(b) we show the systematic errors of each
individual $V_0$, as determined from the fit. The mean systematic
error equal to 0.86\,mV is shown by the horizontal line.
The random error in the mean $V_0$ averaged over all separations
is equal to 0.04\,mV. This finally leads to a
 mean $V_0=-27.4\pm 0.9\,$mV where the total error of 0.9\,mV
 is determined at a 67\% confidence level.
 As can be seen from Fig.~5(a), the mean $V_0$ is independent of
separation over the entire measurement range.
To check this observation, we have performed the best fit of $V_0$
to the straight line [see Fig.~5(a)] leaving its slope as
a free parameter. It was found that the slope is
$0.000012\pm 0.000255\,$mV/nm, i.e.,
the independence of $V_0$ on separation was confirmed.
Note that the larger spread of the individual $V_0$
at larger separations is caused by the smaller values of the total
force measured. This, however, does not influence the systematic
errors in the determination of $V_0$ from the fit [see Fig.~5(b)].
The total error in the mean value of $V_0$ does
not influence the systematic
error in the measurement scheme with fixed $V_0$ which is mostly
determined by instrumental noise (see Sec.~IVA).

The observed independence of $V_0$ on $a$ should be considered in
connection with the problem of patch potentials discussed in Sec.~IIIa.
According to our discussion, the patches due to different crystal
orientations of the polycrystal sample, surface contaminants and dust
may lead to different dependences of $V_0$ on $a$.
Specifically, our simulations show that $V_0$ depends only slightly on
$a$ for patch sizes smaller than the separation distance.
This is the case of patches due to different crystal orientations.
The influence of this type of patches is automatically taken into
account in our measurements either in the value of subtracted
electric force or in the value of applied compensating voltage
(see Sec.~IV).
{}From our simulations it also follows that $V_0$ depends heavily on $a$
for patch sizes of order or larger than the separation distance.
According to Fig.~4, in our experiment $V_0$ remains constant up to
$a=750\,$nm. Thus, the existence on our plate of patches of more
than
2000\,nm size discussed in the literature\cite{49} is incompatible with
our measurement data for $V_0$.
As was mentioned in the introductory part of Sec.~III, the
dependence of $V_0$ on separation was observed in different
experiments on the Casimir force (see, e.g.,
Refs.\cite{39,40a,40,45,47,53a}). It might be caused for different
reasons including the mechanical drift considered above.
In each case the specific reason can be only determined from
a complete analysis of the setup and all the details of that
particular experiment.

The next step was to determine the separation distance at closest approach
$z_0$ and the coefficient $C$ in Eq.~(\ref{eq5}).
As was explained above, these parameters
can be found from the dependence of the parabola curvature
$\beta$ on
distance $a$. The corresponding theoretical expression for parabola
curvature was fit to the measured data for $\beta$ as function of
the separation distance.
A least $\chi^2$-procedure was used in the fitting and
the best values of $z_0$ and $C$ were
obtained. The fitting procedure was repeated by keeping the start point
fixed at the closest separation, while the end point
$z_{\rm end}$ measured from the closest separation was varied from 750\,nm
to 50\,nm. In Fig.~6(a) the $z_0$ so determined is shown as a function
of the end point used in the fit. The systematic errors in the
determination of $z_0$ from the fit vary between 0.36\,nm and
0.48\,nm.
In plotting Fig.~6(a) we have included the correction to the mechanical
drift of separations, as described above.
The values of $z_0$
are seen to be independent on $z_{\rm end}$
indicating the absence of  errors
resulting from $z_{\rm piezo}$ calibration.
Similarly the value of the coefficient $C$ was
also extracted by fitting the
$\beta$-curve as a function of separation to theoretical
expression.
The results obtained after the inclusion
of the correction due to mechanical
drift are shown in Fig.~6(b).
Here, the systematic errors vary between 0.13\,kHz\,m/N and
0.19\,kHz\,m/N.
The independence on $z_{\rm end}$
again indicates the absence of  errors
resulting from $z_{\rm piezo}$ calibration. The mean
values of the calibration parameters obtained are
$z_0=195.9\pm 0.4\,$nm and $C=68.3\pm 0.16\,$kHz\,m/N.
The indicated total errors are mostly determined by the
systematic errors in the fit.
These allow determination of absolute separation between the sphere
and the plate and conversion
of the frequency shift signal  to the
gradient of the total force.

\section{Measurement results for the gradient of the Casimir force}

In this section we present the data obtained for the gradient of the
Casimir force as a function of separation and determine the random,
systematic and total experimental errors. The data for the gradient
of the Casimir force are obtained in two ways: with applied
compensating voltage and with different applied voltages with subsequent
subtraction of the gradient of electric force. The obtained results
are compared with the measured in an earlier experiment.\cite{19,20}

\subsection{Measurement scheme with applied compensating voltage}

To compensate the residual potential difference between the sphere and
the plate one should apply to the plate the potential $V_i=V_0$.
Then the electric force vanishes and from Eq.~(\ref{eq5}) we obtain the
gradient of the Casimir force
\begin{equation}
F^{\prime}(a)=\frac{\partial F(a)}{\partial a}=
-\frac{1}{C}\Delta\omega.
\label{eq6}
\end{equation}
\noindent
The dependence of the Casimir force gradient on separation was measured
40 times. The mean values of the force gradient with a step of one
nanometer are shown as black dots in Fig.~7 and over a more narrow
separation region in the inset to Fig.~7. In the same figure all 40
individual measured values of the Casimir force gradient are shown as
grey dots with a step of 5\,nm (in the inset a step size of 1\,nm
is shown).

The statistical properties of the experimental data measured are
characterized by the histograms presented in Figs.~8(a) for $a=235\,$nm
and 8(b) for $a=275\,$nm. The histograms are described by Gaussian
distributions with the standard deviations equal to
$\sigma_{F^{\prime}}=0.89\,\mu$N/m [Fig.~8(a)] and
$\sigma_{F^{\prime}}=0.87\,\mu$N/m [Fig.~8(b)].
The values of the respective mean gradients of the Casimir force are
${F^{\prime}}=73.58\,\mu$N/m [Fig.~8(a)] and
${F^{\prime}}=41.07\,\mu$N/m [Fig.~8(b)].
The solid and dashed vertical lines are the theoretical predictions
from the plasma and Drude model approaches, respectively
(see Sec.~V for a discussion).

Now we discuss the results of the error analysis.
The random error in the gradient of the Casimir force calculated from 40
repetitions at a 67\% confidence level is shown as the short-dashed line
in Fig.~9 (note that all the experimental errors here and below
are in fact determined with a step of 1\,nm).
The systematic error in the measured gradient is determined by
the instrumental noise including the background noise level, and by the
errors in calibration. In Fig.~9 the systematic error determined at a
67\% confidence level is shown by the long-dashed line.
The solid line in Fig.~9 demonstrates the total experimental error
obtained by adding in quadrature the random and systematic errors.
As can be seen in Fig.~9, all errors do not depend on separation,
as it usually occurs\cite{30} in measurements of the Casimir force by
means of an AFM with applied compensating voltage at separations
above 200\,nm.
The systematic error due to the instrumental noise is
dominant and mostly determines the
value of the total experimental error. The values of the mean measured force
gradients at different separations (first column) together with the total
experimental errors are shown in the second column of Table I.
As can be seen from this table, the relative total experimental error
takes the minimum value of 0.69\% at $a=236\,$nm, and then increases to
0.85\%, 1.7\% 3.0\%, 4.9\%, and 11.6\% at separations $a=250$, 300, 350,
400, and 500\,nm, respectively. At $a=746\,$nm the relative total
experimental error reaches a value of 47\%.

It is of crucial importance to compare the gradients of the Casimir
force measured here by means of the dynamic AFM with the results
of a
previous precision experiment\cite{19,20} performed by means of micromachined
oscillator in Indianapolis University --- Purdue University Indianapolis
(IUPUI).
In Fig.~10(a,b) the magnitudes of the mean Casimir pressure between
two Au-coated
plates determined in Refs.~\cite{19,20} are shown by (a) black and (b)
white lines over different separation regions. In accordance with the PFA,
the magnitudes of the Casimir pressure measured here can be obtained from
the force gradient as
\begin{equation}
|P(a)|=\frac{1}{2\pi R}F^{\prime}(a).
\label{eq7}
\end{equation}
\noindent
In Fig.~10(a,b) the mean pressure magnitudes measured by us are shown as
crosses. The arms of the crosses are determined by the absolute errors in
the measurement of separations and force gradients (the latter are given in
Fig.~9). Note that the error in separation distances determined at a 67\%
confidence level is approximately equal to the error in the point of the
closest approach between the sphere and the plate,
$\Delta a=\Delta z_0=0.4\,$nm.

As can be seen in Fig.~10(a,b), the magnitudes of the mean Casimir pressures
measured by us are in excellent agreement with the experimental results
obtained previously.\cite{19,20}
What's more in most of cases the centers of our experimental crosses are
found in closest proximity to the magnitudes of mean experimental
pressures measured in the IUPUI. To quantify this statement, in Table~I
(column 4) we present the gradients of the Casimir force which are
obtained by multiplication of the pressure magnitudes $|P|$ measured
in the experiment with micromachined oscillator\cite{19,20} by $2\pi R$,
where $R$ is the radius of our sphere.
For convenience in comparison, the total experimental errors are
indicated at the same 67\% confidence level as in our
measurements.
{}From the comparison of columns 2 and 4 in Table~I it
can be seen that
all differences between the respective gradients are in the
limits of the total experimental errors in each experiment.

\subsection{Measurement scheme with different applied voltages}

Now we consider another experimental approach to measuring the gradients
of the Casimir force in sphere-plate geometry, the same as was used to
perform electrostatic calibrations. In this approach different voltages
$V_i$ are applied to the plate while the sphere remains grounded and
the gradient of the total force (electrostatic plus Casimir) is measured.
Then the
gradient of the Casimir force is obtained  from Eq.~(\ref{eq5}) as
\begin{equation}
F^{\prime}(a)=
-\frac{1}{C}\Delta\omega-\frac{\partial X(a,R)}{\partial a}
(V_i-V_0)^2.
\label{eq8}
\end{equation}
\noindent
The dependence of the Casimir force gradient on separation was measured
4 times with 11 applied voltages leading to 44 force-distance curves.
The mean values of the Casimir force gradient with a step of one
nanometer are shown as black dots in Fig.~11(a) and over a more narrow
separation region in the inset. All 44
individual values of the Casimir force gradient are shown as
grey dots with the step of 5\,nm (1\,nm in an inset).
It can be seen that Fig.~11(a) is very similar to Fig.~7 where the
measured gradient of the Casimir force was obtained using another
procedure.

The error analysis in this case is a little different than performed before.
Specifically the random error calculated from 44 repetitions at a 67\%
confidence level is shown by the short-dashed line in Fig.~11(b).
{}From the comparison with Fig.~9 it is seen that in the measurement
scheme with different applied voltages the random error is slightly
smaller. In addition to the two sources of systematic errors discussed
in Sec.~IVA, we now have one more systematic error in the gradient
of electrostatic force subtracted in accordance to Eq.~(\ref{eq8}).
As a result, the systematic error in the gradients of the Casimir force
shown by the long-dashed line in Fig.~11(b) depends on separation.
The total experimental error determined at a 67\% confidence level is
shown by the solid line in Fig.~11(b). At short separations the total
error is slightly larger and at large separations slightly smaller than in
the measurement scheme with applied compensating voltage.
Specifically, at $a=236\,$nm it is equal to 0.75\% and
at $a=500\,$nm to 11.3\%.

We present the values of mean gradients of the Casimir force measured
with different applied voltages at different separations in column 3
of Table~I together with their total experimental errors.
{}From the comparison of column 3 with column 2 it can be seen that
the gradients of the Casimir force measured with different applied
voltages and with the compensating voltage are in very good mutual agreement.
The differences between the values in columns 2 and 3 calculated at any
separation are significantly smaller than the total experimental errors
indicated in Table~I. This confirms the fact that our error analysis is
conservative and the errors are overestimated giving high confidence
to our conclusions with respect to the comparison with theory (see Sec.~V).
In a similar way, the comparison between columns 3 and 4  also
demonstrates a very good agreement between our data and the results of
IUPUI experiment\cite{19,20} within the limits much below allowed ones,
as determined by the absolute errors.

\section{Comparison between experiment and theory}

Now  we compare the experimental data for the gradient of the Casimir
force between the sphere and the plate with the predictions of the
Lifshitz theory. In doing so we adapt the classical Lifshitz formula
for two parallel plates to the sphere-plate geometry using the PFA and
take into account recently calculated corrections to this approximate
method.\cite{54,55}
(The corrections computed previously\cite{55a,55b} cannot be used in
this experiment because they are found for much larger values of
$a/R$.)
In the framework of the PFA, the Lifshitz-type
formula for the gradient of the Casimir force between a sphere and
a plate takes the form
\begin{equation}
F_{\rm PFA}^{\prime}(a,T)=2k_BTR
\sum_{l=0}^{\infty}{\vphantom{\sum}}^{\prime}
\int_{0}^{\infty}q_lk_{\bot}dk_{\bot}\sum_{\alpha}
\frac{r_{\alpha}^2}{e^{2q_la}-r_{\alpha}^2}.
\label{eq9}
\end{equation}
\noindent
Here, $k_B$ is the Boltzmann constant and $T=300\,$K is the
laboratory temperature (we restore this argument, omitted above,
in theoretical equations). The quantity $k_{\bot}$ is the projection
of the wave vector on a plate, $q_l^2=k_{\bot}^2+\xi_l^2/c^2$,
and $\xi_l=2\pi k_BTl/\hbar$ with $l=0,\,1,\,2,\,\ldots$
are the Matsubara frequencies. The prime following the summation
sign multiplies the term with $l=0$ by 1/2 and $\alpha={\rm TM,\,TE}$
denotes the transverse magnetic and transverse electric polarizations
of the electromagnetic field. The reflection coefficients $r_{\alpha}$
calculated along the imaginary frequency axis are given by
\begin{eqnarray}
&&
r_{\rm TM}\equiv r_{\rm TM}(i\xi_l,k_{\bot})=
\frac{\varepsilon(i\xi_l)q_l-k_l}{\varepsilon(i\xi_l)q_l+k_l},
\nonumber \\
&&
r_{\rm TE}\equiv r_{\rm TE}(i\xi_l,k_{\bot})=
\frac{q_l-k_l}{q_l+k_l},
\nonumber \\
&&
k_l=\left[k_{\bot}^2+\varepsilon(i\xi_l)\frac{\xi_l^2}{c^2}
\right]^{1/2},
\label{eq10}
\end{eqnarray}
\noindent
where $\varepsilon(i\xi_l)$ is the dielectric permittivity of
boundary materials at the imaginary frequencies.

Computations of $F^{\prime}(a,T)$ using Eq.~(\ref{eq9}) were
performed with the two models of the dielectric permittivity
of Au called in the literature the Drude model approach and
the plasma model approach.\cite{3,30} In the Drude model
approach, the tabulated optical data\cite{56} for the imaginary
part of dielectric permittivity of Au are used. They are
extrapolated to lower frequencies by means of the imaginary part
of the Drude model with the plasma frequency $\omega_p=9.0\,$eV
and the relaxation parameter $\gamma=0.035\,$eV.
Recently it was shown\cite{57} that $\varepsilon(i\xi_l)$
obtained with this extrapolation is in excellent agreement with
$\varepsilon(i\xi_l)$ obtained with the help of the weighted
Kramers-Kronig relations from the measured tabulated data.
In the plasma model approach, the same optical data with the
contribution of free charge carriers subtracted
are extrapolated to lower
frequencies by means of the simple plasma model with the same value
of the plasma frequency for Au.

The correction to the approximate expression (\ref{eq9}) was
recently calculated in the framework of the exact theory.\cite{54,55}
The exact force gradient between the sphere of large radius and the
plate was represented in the form
\begin{equation}
F^{\prime}(a,T)=F_{\rm PFA}^{\prime}(a,T)\left[1+
\theta(a,T)\frac{a}{R}+o\left(\frac{a}{R}\right)\right],
\label{eq11}
\end{equation}
\noindent
where the quantity $\theta(a,T)$ was calculated. Specifically,
for ideal metal bodies at $T=0$ it was found\cite{54,55,57a}
\begin{equation}
\theta(a,T)=\frac{1}{9}-\frac{20}{3\pi^2}=-0.564.
\label{eq12}
\end{equation}
\noindent
General expressions for $\theta(a,T)$ were also provided for real
material bodies described by the frequency-dependent dielectric
permittivity at nonzero temperature. In the framework of the
Drude model approach, as described above, the quantity $\theta(a,T)$
was computed\cite{55} as a function of separation at $T=300\,$K.
It was found that $\theta(a,T)$ increases monotonically from --0.438
to --0.329 when $a$ increases from 222\,nm to 642\,nm.
This means that the error from using the PFA, which was taken equal
to $a/R$ in the analysis of previous
experiments\cite{17,18,19,20,25,26,34} was in fact overestimated.
Therefore that analysis should be characterized as highly
conservative.

To compare experiment with theory, one should also take into account
the surface roughness. The roughness profiles on both surfaces of
sphere and plate were investigated using an AFM.
The root-mean-square roughness on the sphere and the plate was found
to be $\delta_s=2.0\,$nm and $\delta_p=1.8\,$nm, respectively.
We have averaged the computed gradients of the Casimir force (\ref{eq11})
to calculate the force gradient between rough surfaces
$F_{\rm theor}^{\prime}$ (the method of geometrical averaging\cite{3,30}).
At separations considered in this experiment ($a\geq 235\,$nm) the same
results for $F_{\rm theor}^{\prime}$ were obtained after the multiplication
of Eq.~(\ref{eq11}) by the factor
\begin{equation}
\eta_R(a)=1+10\frac{\delta_s^2+\delta_p^2}{a^2}+
105\frac{(\delta_s^2+\delta_p^2)^2}{a^4},
\label{eq13}
\end{equation}
\noindent
i.e., using the multiplicative approach.\cite{3,30}
This is explained by the fact that at such large separations and small
roughness the influence of roughness on force gradients is very small.
Thus, at the shortest separation $a=235\,$nm it contributes only 0.13\%
of the force gradient. The role of surface roughness further decreases
with the increase of separation between the surfaces.

The comparison of the experimental data obtained with applied
compensating voltage to the plate (Sec.~IVA) with theory is shown in
Fig.~12(a-d) over different separation regions.
The experimental data are shown as crosses with the total experimental
errors determined at a 67\% confidence level. The exact theoretical
results for $F_{\rm theor}^{\prime}$ computed using the Drude model
approach, as explained above, are shown by the black bands. The widths
of these bands are determined by the error in the sphere radius and
by the errors in the optical data of Au defined by the number of
significant figures in the tables.\cite{56}
Recall that the use of alternative optical data\cite{58} and
respective values of $\omega_p\leq 8\,$eV makes $F_{\rm theor}^{\prime}$
much smaller and, thus, further increases descrepancy between
experiment and theory. Furthermore, for such optical data a significant
disagreement was found\cite{57} between the dielectric permittivities
obtained by the extrapolation using the Drude model and by the
weighted Kramers-Kronig relations.
By the grey bands in Fig.~12(a-d) we present the theoretical results
for $F_{\rm theor}^{\prime}$ as a function of separation found using
the plasma model approach. They are computed by Eq.~(\ref{eq9})
multiplied by the factor (\ref{eq13}) to take into account
the surface roughness. The error arising from the use of the PFA
is included in the widths of grey bands.
These widths take into account that the correction due to
inaccuracy of the PFA is between $-0.564a/R$ and zero
(it was shown\cite{55,55a,55b} that for real metals at $T\neq 0$
the magnitude of the main correction to the PFA is smaller
than for ideal metal).

{}From Fig.~12(a-d) it can be seen that theoretical predictions
obtained using the plasma model approach are in excellent agreement
with the data over the entire range of separations.
As to the predictions of the Drude model approach, they are excluded
by the measurement data over the wide separation region from 235 to
420\,nm. At larger separation distances the vertical arms of the
crosses only touch the theoretical band predicted by the Drude model
approach, whereas the centers of crosses are still far away from the
theoretical predictions. Thus, the experimental data obtained with
applied compensating voltage are consistent with the predictions of the
plasma model approach and exclude the Drude model approach.
This is in accordance with the results obtained previously using another
experimental technique.\cite{17,18,19,20}

We now compare with theory the experimental gradients of the Casimir
force measured with different applied voltages to the plate (see
Sec.~IVB). In this case the results of the comparison are shown in
Fig.~13(a-d), where the experimental data are indicated as crosses.
As in Fig.~12, the black bands are computed using the exact theory in
the framework of the Drude model approach.
The grey bands are computed using the PFA and the plasma model approach.
The error in the plasma model approach is included in the widths of grey
bands.
The results of the comparison between experiment and theory are the same
as in Fig.~12. The plasma model approach is found in excellent agreement
with the experimental data over the entire measurement range (shown in
Fig.~13 and also at larger separations). The Drude model approach is
excluded by the data over the separation region from 235 to 420\,nm at
a 67\% confidence level. At larger separations the vertical arms of the
crosses only touch the black theoretical bands whereas the centers of
the crosses continue to belong to the grey bands.

We emphasize that the use of the exact theory in computations of
the black bands (the Drude model approach) does not
influence the obtained conclusions. Although the correction to the PFA
result in Eq.~(\ref{eq11}) is negative and slightly increases the
deviation between the data and the black bands in Figs.~12 and 13,
the separation range where the Drude model approach is excluded
(from 235 to 420\,nm) remains the same irrespective of whether the
exact theory or the PFA is used.

\section{Nonlinear effects in dynamic technique}

As was noted in Sec.~II, in the dynamic technique,
when the cantilever is oscillating, the separation distance
between the sphere and the plate is varied harmonically in time
\begin{equation}
a(t)=a+A_z\cos\omega_rt,
\label{eq14}
\end{equation}
\noindent
where $\omega_r$ is the resonant frequency of the cantilever under the
influence of the Casimir force and $A_z$ is the oscillation amplitude
which was chosen to be less than 10\,nm. It was supposed that under this
condition at  separations under consideration our oscillation system
belongs to the linear regime where the shift of the natural frequency
is given by Eq.~(\ref{eq2}). Here, we derive the analytic expression for
the frequency shift in the nonlinear regime of an oscillator and
determine the application region of Eq.~(\ref{eq2}).

The expressions for the shift of frequency of a nonlinear oscillator under
the influence of the Casimir (Casimir-Polder) force were found
perturbatively for the micromachined oscillator\cite{4,5} and exactly
for the Bose-Einstein condensate cloud above a plate.\cite{59}
The techniques involving shifts of the resonant frequency under
the influence of an external force was discussed for the purpose
of precise force measurements using different
setups.\cite{40b,63a}
The exact expression\cite{59} adapted to the case of an AFM with attached
sphere in the nonlinear regime is given by
\begin{equation}
\omega_r^2-\omega_0^2=-\frac{\omega_r\omega_0^2}{\pi kA_z}
\int_{0}^{2\pi/\omega_r}\!\!\!\!dt\cos(\omega_rt)
F[a+A_z\cos(\omega_rt),T],
\label{eq15}
\end{equation}
\noindent
where $F$ is the Casimir force acting between the sphere and the plate.
Note that a similar equation was used\cite{60} to investigate the
nonlinear regime for a micromachined oscillator with attached cylinder
interacting with a plate. Here we consider the measurement scheme with the
applied compensating voltage when only the Casimir force causes the
frequency shift of the oscillator.
As was discussed in Sec.~V, under the condition $a\ll R$ one can put
with sufficient precision
\begin{eqnarray}
&&
F(z,T)=F_{\rm PFA}(z,T)=k_BTR
\sum_{l=0}^{\infty}{\vphantom{\sum}}^{\prime}
\int_{0}^{\infty}k_{\bot}dk_{\bot}
\nonumber \\
&&~~~~~~~
\times
\sum_{\alpha}\ln\left(1-r_{\alpha}^2e^{-2q_lz}\right).
\label{eq16}
\end{eqnarray}
\noindent
Substituting Eq.~(\ref{eq16}) into Eq.~(\ref{eq15}) and expanding
the logarithms into power series, one obtains
\begin{eqnarray}
&&
\omega_r^2-\omega_0^2=\frac{\omega_r\omega_0^2}{\pi kA_z}k_BTR
\int_{0}^{2\pi/\omega_r}\!\!\!\!dt\cos(\omega_rt)
\label{17} \\
&&~\times
\sum_{l=0}^{\infty}{\vphantom{\sum}}^{\prime}
\int_{0}^{\infty}\!\!\!k_{\bot}dk_{\bot}
\sum_{n=1}^{\infty}\frac{r_{\rm TM}^{2n}+r_{\rm TE}^{2n}}{n}
e^{-2q_ln[a+A_z\cos(\omega_rt)]}.
\nonumber
\end{eqnarray}
\noindent
Changing the order of summations and integrations and introducing the new
variable $x=\omega_rt$, we arrive at
\begin{eqnarray}
&&
\omega_r^2-\omega_0^2=\frac{\omega_0^2}{\pi kA_z}k_BTR
\sum_{l=0}^{\infty}{\vphantom{\sum}}^{\prime}
\sum_{n=1}^{\infty}\frac{1}{n}
\int_{0}^{\infty}k_{\bot}dk_{\bot}
\nonumber \\
&&\times
(r_{\rm TM}^{2n}+r_{\rm TE}^{2n})
e^{-2naq_l}\!\!
\int_{0}^{2\pi}\!\!\!\!\!dx\cos{x}e^{-2nq_lA_z\cos{x}}.
\label{eq18}
\end{eqnarray}
\noindent
The latter integral can be calculated explicitly\cite{61}
\begin{equation}
\int_{0}^{2\pi}dx\cos{x}e^{-2nq_lA_z\cos{x}}=-2\pi I_1(2nq_lA_z),
\label{eq19}
\end{equation}
\noindent
where $I_n(z)$ is the Bessel function of imaginary argument.

Taking into account that the frequency shift under the influence of the
Casimir force is small in comparison with the resonant frequency, it holds
\begin{equation}
\omega_r^2-\omega_0^2=(\omega_r-\omega_0)(\omega_r+\omega_0)
\approx 2\omega_0(\omega_r-\omega_0).
\label{eq20}
\end{equation}
\noindent
Then Eq.~(\ref{eq18}) can be rewritten as
\begin{eqnarray}
&&
\omega_r-\omega_0=-\frac{\omega_0}{kA_z}k_BTR
\sum_{l=0}^{\infty}{\vphantom{\sum}}^{\prime}
\sum_{n=1}^{\infty}\frac{1}{n}
\int_{0}^{\infty}k_{\bot}dk_{\bot}
\nonumber \\
&&~~~~~~~~\times
(r_{\rm TM}^{2n}+r_{\rm TE}^{2n})
e^{-2naq_l}I_1(2nq_lA_z).
\label{eq21}
\end{eqnarray}
\noindent
In terms of dimensionless variable $y=2aq_l$ Eq.~(\ref{eq21}) takes
the form
\begin{eqnarray}
&&
\omega_r-\omega_0=-\frac{\omega_0}{4a^2 kA_z}k_BTR
\sum_{l=0}^{\infty}{\vphantom{\sum}}^{\prime}
\sum_{n=1}^{\infty}\frac{1}{n}
\nonumber \\
&&~~\times
\int_{\zeta_l}^{\infty}ydy
(r_{\rm TM}^{2n}+r_{\rm TE}^{2n})
e^{-ny}I_1\left(\frac{A_z}{a}ny\right),
\label{eq22}
\end{eqnarray}
\noindent
where $\zeta_l=2a\xi_l/c$ is the dimensionless Matsubara frequency.
This is the final analytic expression for the frequency shift of a
cantilever in the nonlinear regime.

Let us compare Eq.~(\ref{eq22}) with Eq.~(\ref{eq2}) and determine the
application region of the latter. For this purpose we represent the
Bessel function as power series\cite{61}
\begin{equation}
I_1(z)=\frac{z}{2}+\frac{z^3}{16}+O(z^5)
\label{eq23}
\end{equation}
\noindent
and substitute the first two terms into Eq.~(\ref{eq22}):
\begin{eqnarray}
&&
\omega_r-\omega_0=-\frac{\omega_0}{8a^3 k}k_BTR
\sum_{l=0}^{\infty}{\vphantom{\sum}}^{\prime}
\sum_{n=1}^{\infty}
\int_{\zeta_l}^{\infty}y^2dy
\nonumber \\
&&~~~
\times (r_{\rm TM}^{2n}+r_{\rm TE}^{2n})e^{-ny}
\label{eq24}\\
&&~
-\frac{\omega_0A_z^2}{64a^5 k}k_BTR
\sum_{l=0}^{\infty}{\vphantom{\sum}}^{\prime}
\sum_{n=1}^{\infty}n^2
\int_{\zeta_l}^{\infty}\!\!\!\!y^4dy
(r_{\rm TM}^{2n}+r_{\rm TE}^{2n})e^{-ny}.
\nonumber
\end{eqnarray}
\noindent
The sum in $n$ in the first term on the right-hand side of Eq.~(\ref{eq24})
is calculated as
\begin{equation}
\sum_{n=1}^{\infty}(r_{\rm TM}^2+r_{\rm TE}^2)e^{-ny}=
\frac{r_{\rm TM}^2}{e^y-r_{\rm TM}^2}
+
\frac{r_{\rm TE}^2}{e^y-r_{\rm TE}^2}.
\label{eq25}
\end{equation}
\noindent
By comparing Eq.~(\ref{eq24}) and (\ref{eq9}) with account of
Eq.~(\ref{eq25})
and the connection between dimensional and dimensionless variables,
we arrive at
\begin{eqnarray}
&&
\omega_r-\omega_0=-\frac{\omega_0}{2k}
\frac{\partial F_{\rm PFA}(a,T)}{\partial a}
-\frac{\omega_0A_z^2}{64a^5 k}k_BTR
\nonumber\\
&&~~~
\times
\sum_{l=0}^{\infty}{\vphantom{\sum}}^{\prime}
\sum_{n=1}^{\infty}n^2
\int_{\zeta_l}^{\infty}\!\!\!\!y^4dy
(r_{\rm TM}^{2n}+r_{\rm TE}^{2n})e^{-ny}.
\label{eq26}
\end{eqnarray}
\noindent
Furthermore, taking into account that
\begin{eqnarray}
&&
\sum_{n=1}^{\infty}n^2(r_{\rm TM}^2+r_{\rm TE}^2)e^{-ny}=
r_{\rm TM}^2e^{-y}\frac{1+r_{\rm TM}^2e^{-y}}{(1-r_{\rm TM}^2e^{-y})^3}
\nonumber \\
&&~~~~~~~
+
r_{\rm TE}^2e^{-y}\frac{1+r_{\rm TE}^2e^{-y}}{(1-r_{\rm TE}^2e^{-y})^3},
\label{eq27}
\end{eqnarray}
\noindent
we rewrite Eq.~(\ref{eq26}) in the form
\begin{eqnarray}
&&
\omega_r-\omega_0=-\frac{\omega_0}{2k}
\frac{\partial F_{\rm PFA}(a,T)}{\partial a}
-\frac{\omega_0A_z^2}{64a^5 k}k_BTR
\nonumber\\
&&~~~
\times
\sum_{l=0}^{\infty}{\vphantom{\sum}}^{\prime}
\int_{\zeta_l}^{\infty}\!\!\!\!y^4dy
\left[
r_{\rm TM}^{2}e^{-y}
\frac{1+r_{\rm TM}^{2}e^{-y}}{(1-r_{\rm TM}^{2}e^{-y})^3}
\right.
\label{eq28} \\
&&~~~~~~
\left.+
r_{\rm TE}^{2}e^{-y}
\frac{1+r_{\rm TE}^{2}e^{-y}}{(1-r_{\rm TE}^{2}e^{-y})^3}
\right].
\nonumber
\end{eqnarray}
\noindent
Here, the first term on the right-hand side coincides with the
right-hand side of Eq.~(\ref{eq2}) (the linear regime), whereas
the second term describes nonlinear corrections (note that the
total force now coincides with the Casimir force).

One can restrict oneself to the linear regime if the magnitude of
the second term is much smaller than that of the first. Keeping in mind that
the force gradient is connected with the pressure by means of
Eq.~(\ref{eq7}), this condition can be written as
\begin{eqnarray}
&&
|P(a,T)|\gg \frac{k_BTA_z^2}{64a^5}
\sum_{l=0}^{\infty}{\vphantom{\sum}}^{\prime}
\!\!\int_{\zeta_l}^{\infty}\!\!\!\!y^4dy
\label{eq29} \\
&&~~\times
\sum_{\alpha}
\left[
r_{\rm TM}^{2}e^{-y}
\frac{1+r_{\rm TM}^{2}e^{-y}}{(1-r_{\rm TM}^{2}e^{-y})^3}
+
r_{\rm TE}^{2}e^{-y}
\frac{1+r_{\rm TE}^{2}e^{-y}}{(1-r_{\rm TE}^{2}e^{-y})^3}
\right].
\nonumber
\end{eqnarray}
\noindent
We have calculated the quantity in the
right-hand side of Eq.~(\ref{eq29}) for the
parameters of our experimental setup under the condition that
this quantity does
not exceed 1\% of the magnitude of the Casimir pressure.
The obtained maximum
allowed oscillation amplitudes of the cantilever are shown
in Fig.~14 as a function of
separation by the solid and dashed lines for the plasma and Drude
model approaches, respectively. The allowed regions in the plane $(a,A_z)$,
where the contribution of nonlinear effects is less than 1\%, lie beneath
the lines. As an example, at separations 100, 235, 300 and 500\,nm
the oscillation amplitude should not exceed 7.07, 16.1, 20.2, and 32.8\,nm
if computations are performed using the Drude model approach.
If computations are performed using the plasma model approach only
slightly different maximum amplitudes are allowed. They are equal to
7.11, 16.2, 20.5, and 33.4\,nm at the same respective separations.

Note that the use of full Eqs.~(\ref{eq22}) and (\ref{eq28})
opens opportunities
for performing measurements in the  nonlinear regime. In this case
the immediately measured quantity would be the frequency shift to be
compared with theoretical computations using Eqs.~(\ref{eq22}) or
(\ref{eq28}) with different dielectric properties of boundary
surfaces.

\section{Conclusions and Discussion}

In the foregoing we have presented the results of precise measurements
of the gradient of the Casimir force between an Au-coated sphere and
a plate by means of an AFM operated in the dynamic regime.
{}From several modifications of dynamic measurements the most sensitive
frequency shift technique has been employed which has never been used
before in experiments on the Casimir force using an AFM. This was connected
with creation of significantly different setup adapted for dynamic
measurements, use of higher vacuum and hollow glass microspheres of
smaller radius.

Special attention was devoted to electrostatic calibrations of the setup,
i.e., to a problem which created much discussion in previous literature.
We have addressed in much detail both the problem of electrostatic patches
and contaminants on the surface and the problem of dependence of the
calibration parameters on separation between the test bodies.
It is well known that many different models of patches were discussed
in the literature leading to varying predictions of additional electrostatic
forces from large\cite{49} to negligibly small.\cite{62}
To address this problem, we performed numerical simulation of the
electrostatic force due to the patch potentials and have shown that for both
relatively small and large patches the residual potential between the
sphere and the plate would be separation-dependent for patch sizes of order
or larger than the separation. This adds importance to the second problem,
i.e., to the separation-dependence of the calibration parameters.
Experimental investigation of this problem demonstrated that the calibration
parameters are constant if the corrections to mechanical drift are
introduced. Our measurement data for electric forces unequivocally exclude
the predicted\cite{49} large electrostatic force from the polycrystal
structure of Au coatings. On the other hand, the observed independence
of the residual potential difference on separation rejects the
hypothesis\cite{49} of large contaminants on the surface, which could
decrease significantly the enormously large effect of a polycrystal
structure, but, as follows from our simulations,
 lead to the separation-dependent residual potential.
These findings are in line with the fact that the surfaces of the Au-coated
sphere and the plate in our experiment have been subjected to a multistep
cleaning procedure and would be unlikely to have large contaminants.
Thus, our conclusion is that the experimental data are in favor of the
model of patches proposed previously.\cite{62}

It should be stressed that the mean gradient of the Casimir force as
a function of separation has been measured in our experiment in two
independent ways (with applied compensating voltage and with applied
different voltages to the plate with subsequent subtraction of electric
forces). The obtained results were found in excellent agreement in the
limits of total experimental errors. The latter were determined as
combinations of random and systematic errors at a 67\% confidence level.
The mean measured gradients of the Casimir force were converted into the
pressure between two parallel plates and compared with respective results
measured in the most precise experiment performed by means of micromachined
oscillator.\cite{19,20} The mean Casimir pressures determined in both
experiments were found in excellent agreement over the entire measurement
range.

The mean measured gradients of the Casimir force were compared with
theoretical predictions of the Lifshitz theory with no fitting parameters.
In so doing two theoretical approaches proposed in the literature were
used based on the Drude and plasma models of dielectric permittivity.
The contribution of surface roughness was calculated to be less than
0.13\% of the force gradients.
The measured data were shown to be consistent with theoretical results
obtained using the generalized plasma-like model over the entire
measurement range. Theoretical predictions computed using the Drude
model approach were excluded by the data over the separation region from
235 to 420\,nm at a 67\% confidence level with measurements at every
nanometer.
The nonlinear regime of our oscillator was investigated, and the linearity
in the region of used experimental parameters was confirmed.
It is pertinent to note that one experiment alone
performed at a 67\% confidence level would be not enough to
falsify application in Casimir physics of a well tested
and commonly used theoretical model.
In this regard our experiment should be considered as an
additional independent argument to more precise experiments
(up to 99.9\% confidence level) performed using another
experimental technique.\cite{3,18,19,20}

The main result on the exclusion of the Drude model by the data
deserves special discussion. It is common knowledge that
response of metallic materials to real electromagnetic fields is correctly
described by the Drude model, whereas the plasma model is only an
approximation valid in the region of sufficiently high frequencies.
Experiments using a micromachined oscillator\cite{17,18,19,20} demonstrated
that in the Lifshitz theory not the Drude but the plasma model is
supported by the data. Thereafter many attempts were undertaken to rule out
this conclusion. This is the reason why one more experiment,
using an alternative laboratory setup, is highly desirable.
In our experiment, using a dynamic AFM in a frequency modulation technique,
we confirmed the results of previous measurements performed by means
of a micromachined oscillator. We also addressed the
problem of patch potentials, independence of the calibration parameters on
separation and applicability of the linear regime of the dynamic AFM.
In addition, an independent comparison between experiment and theory beyond
the PFA has been made with no fitting parameters. Nevertheless the Drude
model approach was again excluded by the data. One can conclude that
the exclusion of this approach to a theoretical description of the Casimir
interaction between metallic surfaces received a more complete
experimental confirmation.
Keeping in mind similar experiments with semiconductor
and dielectric test bodies discussed in Sec.~I, a thorough analysis of all
the assumptions in the basics of the Lifshitz theory seems pertinent.
\section*{Acknowledgments}
The authors are grateful to G.~Bimonte for providing the
numerical values of a correction beyond the PFA
from Fig.~2 of Ref.\cite{55}.
This work was supported by the  NSF Grant
No.~PHY0970161 (C.-C.C., G.L.K., V.M.M., U.M.), DOE Grant
No.~DEF010204ER46131 (equipment, G.L.K., V.M.M., U.M.)
and DARPA Grant under Contract
No.~S-000354 (A.B., R.C.-G., U.M.).
 G.L.K.\ and V.M.M.\ were also partially
supported by the DFG grant BO\ 1112/21--1.


\begin{figure*}[h]
\vspace*{-1.0cm}
\centerline{\hspace*{-1cm}
\includegraphics{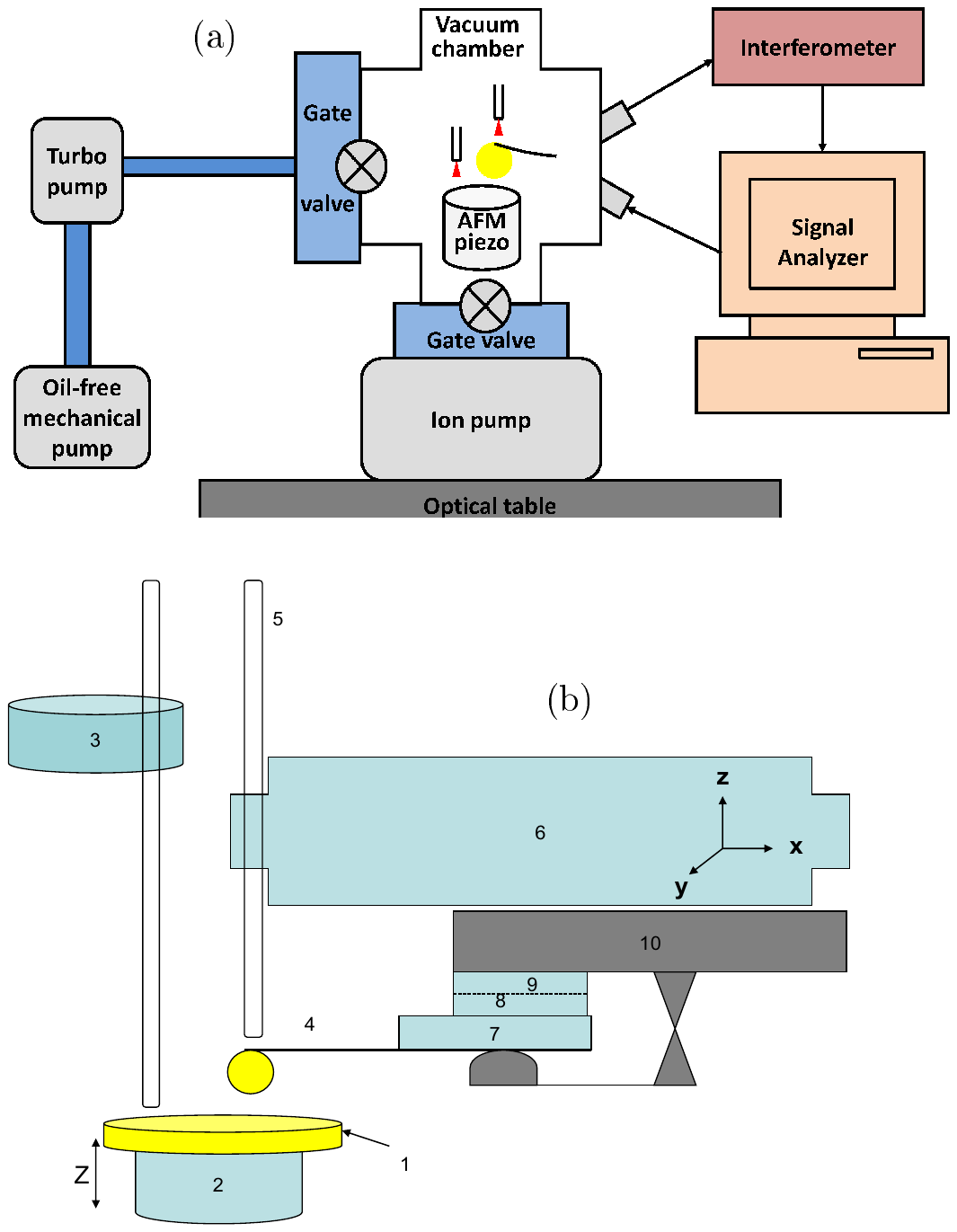}
}
\vspace*{-13.5cm}
\caption{(Color online)
(a) Layout of the vacuum FM-AFM setup used in precision dynamic
measurements of the gradient of the Casimir force.
(b) Schematic of the force measurements microscope.
(1) is a Au plate placed on the AFM piezo (2). (3) is the plate movement interferometer
detection fiber. For monitoring the cantilever (4) oscillations the second interferometer
was used, the detection fiber end is shown as (5). The end was fixed in the fiber holder (6),
which was placed in the XYZ stage and can move in the XYZ direction for adjusting the signal
from cantilever. The cantilever chip (7) was connected to two piezoelectric actuators (8 and 9)
and clutched in the home-made cantilever holder (10) as shown in the picture.
}
\end{figure*}
\begin{figure*}[h]
\vspace*{-5.0cm}
\centerline{\hspace*{-1cm}
\includegraphics{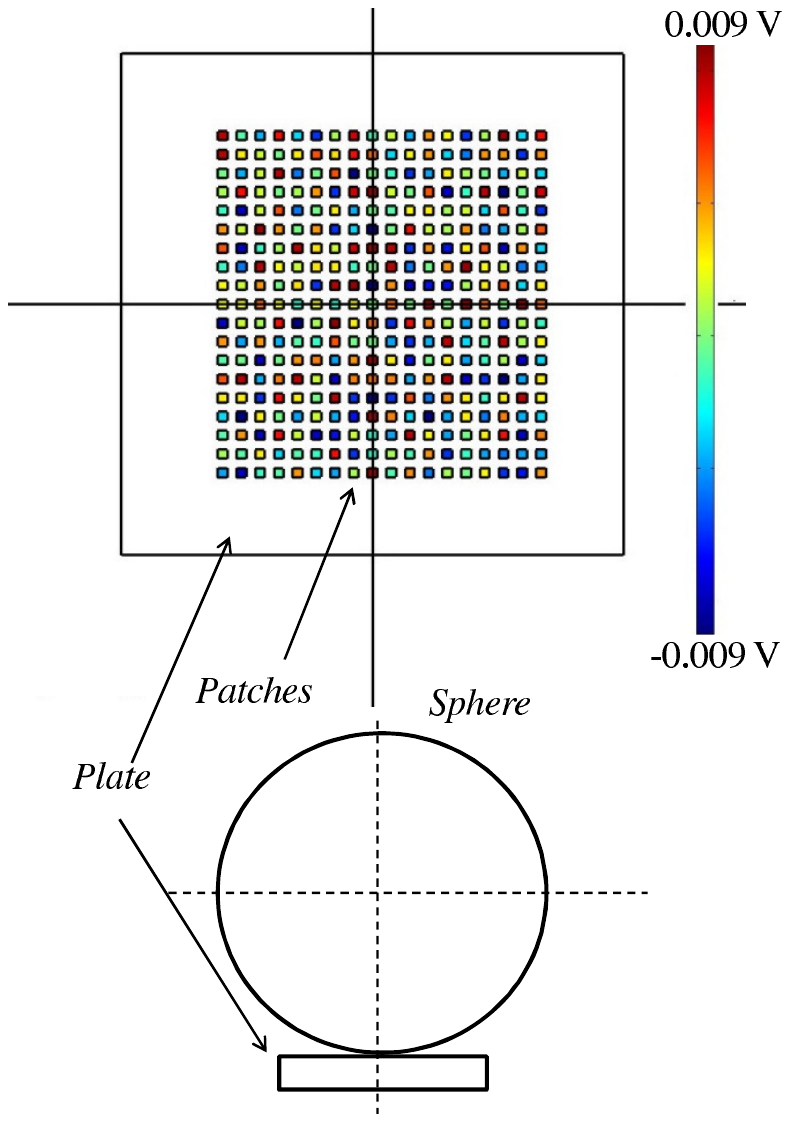}
}
\vspace*{-11.cm}
\caption{(Color online)
The sphere-plate configuration with patches which were put
on the top of a plate.
}
\end{figure*}
\begin{figure*}[h]
\vspace*{-3.cm}
\centerline{\hspace*{-1cm}
\includegraphics{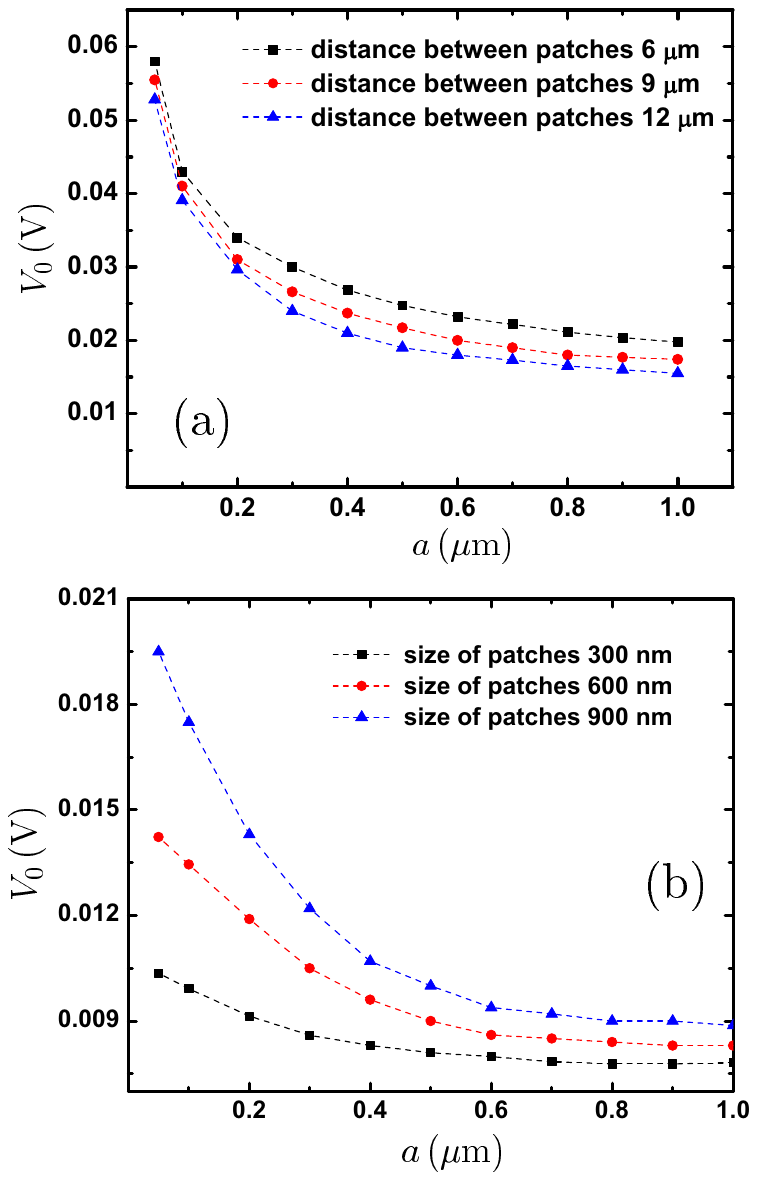}
}
\vspace*{-15.cm}
\caption{(Color online)
The residual potential difference between the sphere and
the plate as a function of separation for (a) patches of fixed
sizes $6\times 6\,\mu\mbox{m}^2$
with different distances between patches and
(b) fixed distances between patches equal to 600\,nm with
different sizes of patches.
}
\end{figure*}
\begin{figure*}[h]
\vspace*{-7.cm}
\centerline{\hspace*{1cm}
\includegraphics{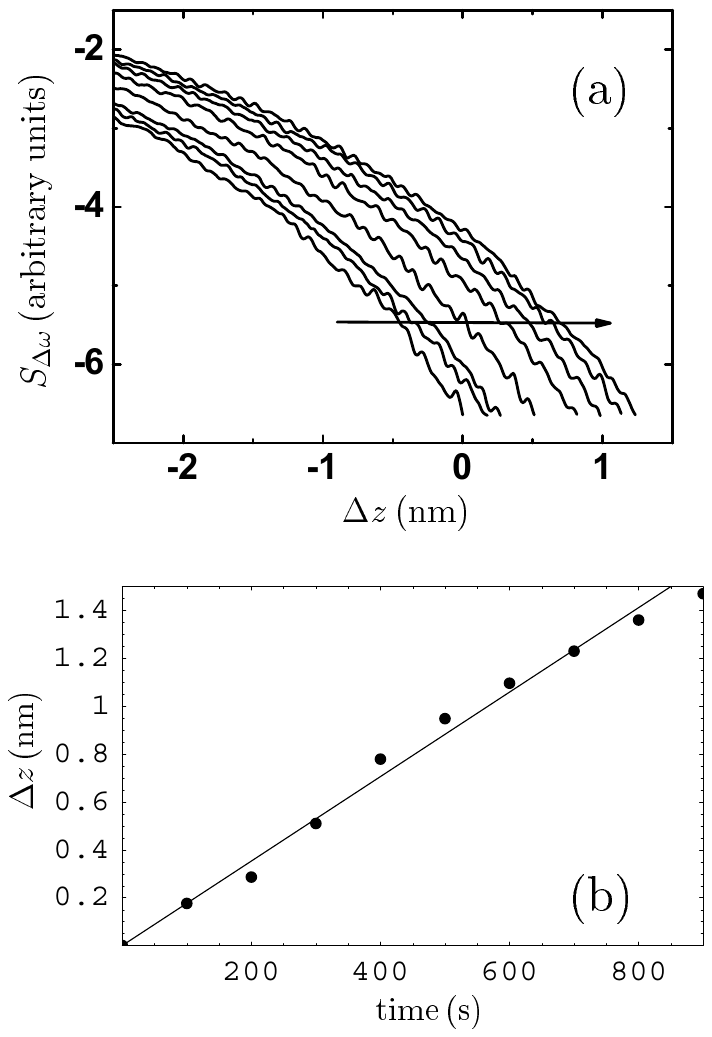}
}
\vspace*{-11.5cm}
\caption{(a) The drift in experimental curves for the frequency
 shift signal as a function of the change in sphere-plate separation
for 8 repetitions of the same applied voltage to the plate.
(b) The change in the plate position at one frequency shift signal
value as
a function of time.
}
\end{figure*}
\begin{figure*}[h]
\vspace*{-4.cm}
\centerline{\hspace*{1cm}
\includegraphics{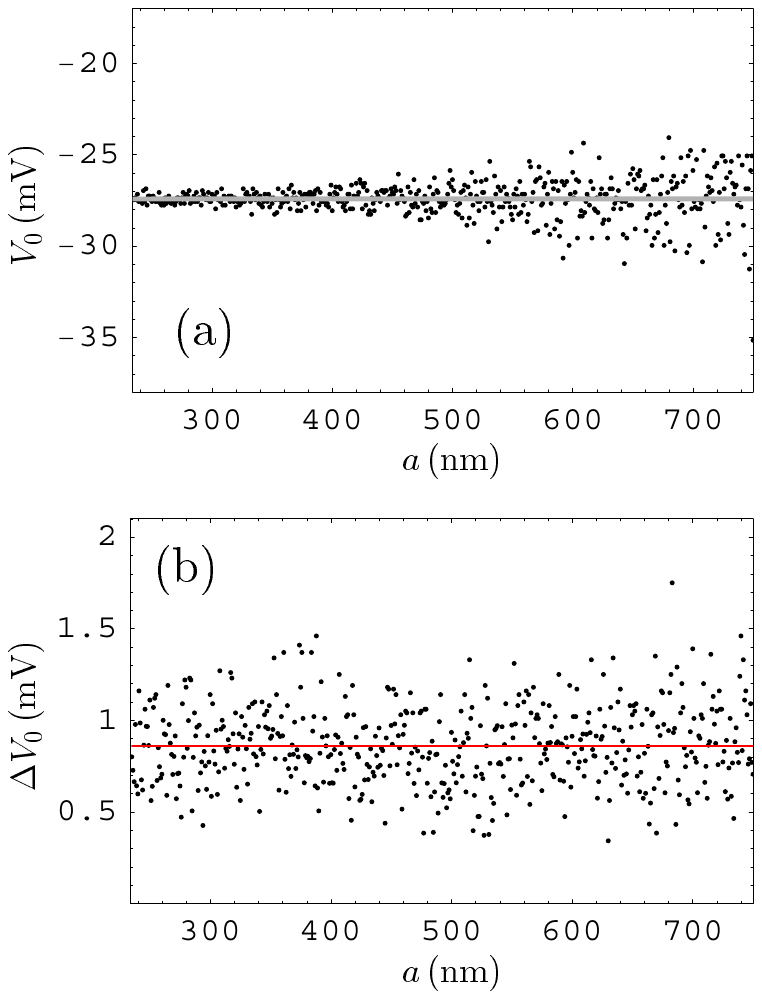}
}
\vspace*{-11.5cm}
\caption{(Color online) (a) The residual potential difference between Au-coated
sphere and plate as a function of separation.
(b) The systematic error of each individual $V_0$, as determined
from the fit, versus separation.
}
\end{figure*}
\begin{figure*}[h]
\vspace*{-8.cm}
\centerline{\hspace*{1cm}
\includegraphics{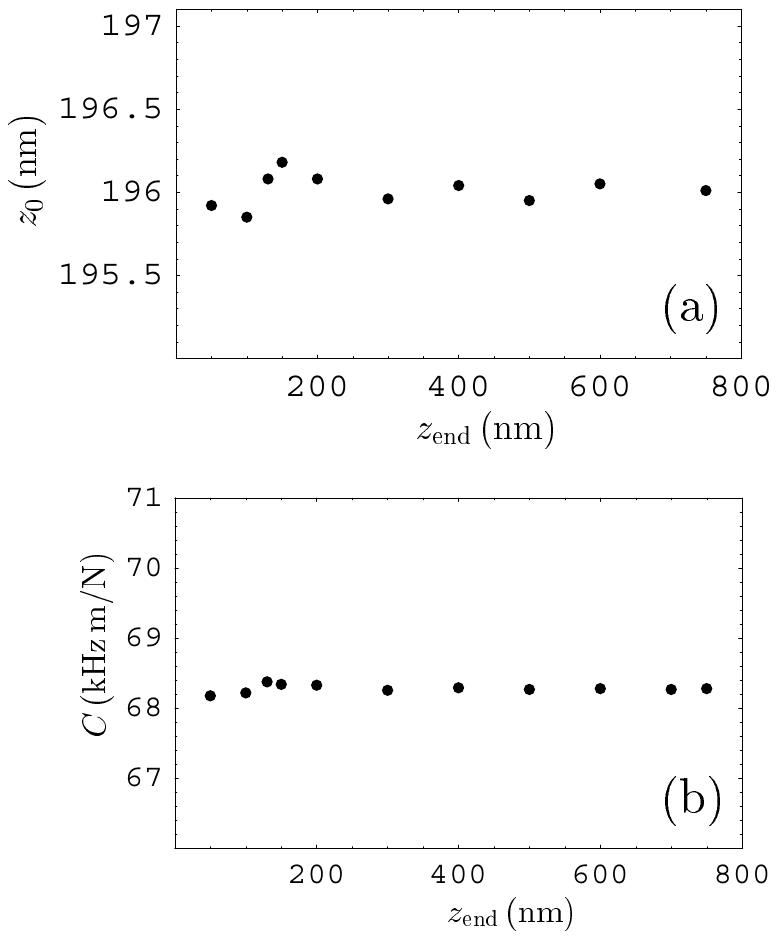}
}
\vspace*{-11.5cm}
\caption{The dependences of (a) the closest sphere-plate separation and
(b) the coefficient $C$ in Eq.~(\ref{eq5}) on the end point of the fit.
}
\end{figure*}
\begin{figure*}[h]
\vspace*{-12.cm}
\centerline{\hspace*{1cm}
\includegraphics{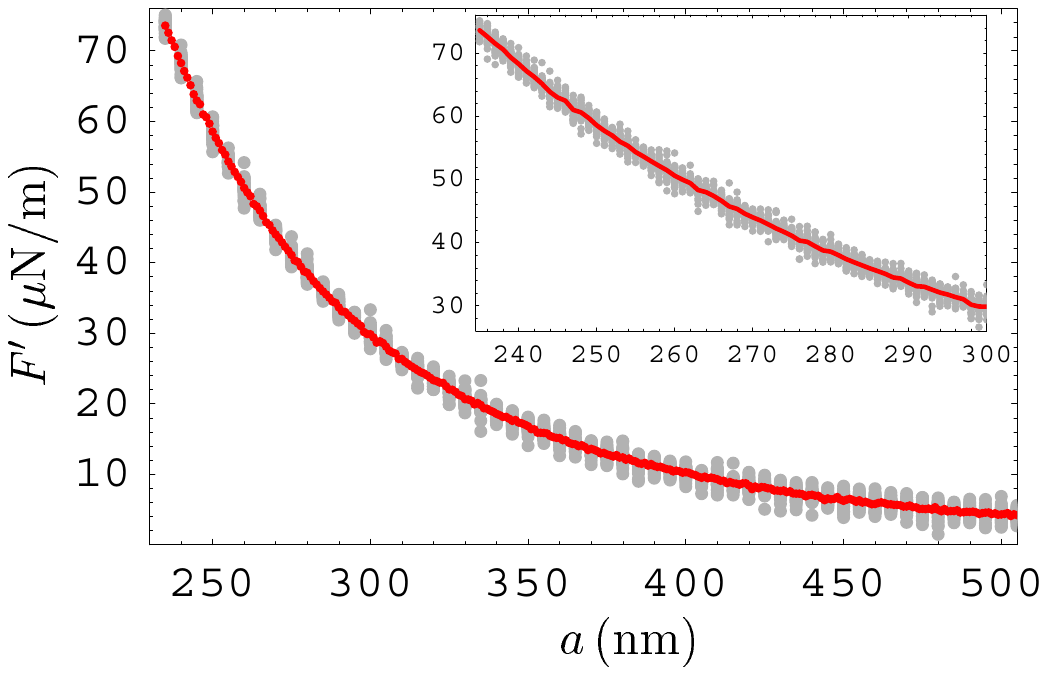}
}
\vspace*{-6.5cm}
\caption{(Color online) Mean measured gradients of the Casimir force
as a function of separation are shown by solid lines.
Grey dots indicate all 40 individual force gradients plotted with a step
of 5\,nm (1\,nm in the inset).
}
\end{figure*}
\begin{figure*}[h]
\vspace*{-5.cm}
\centerline{\hspace*{1cm}
\includegraphics{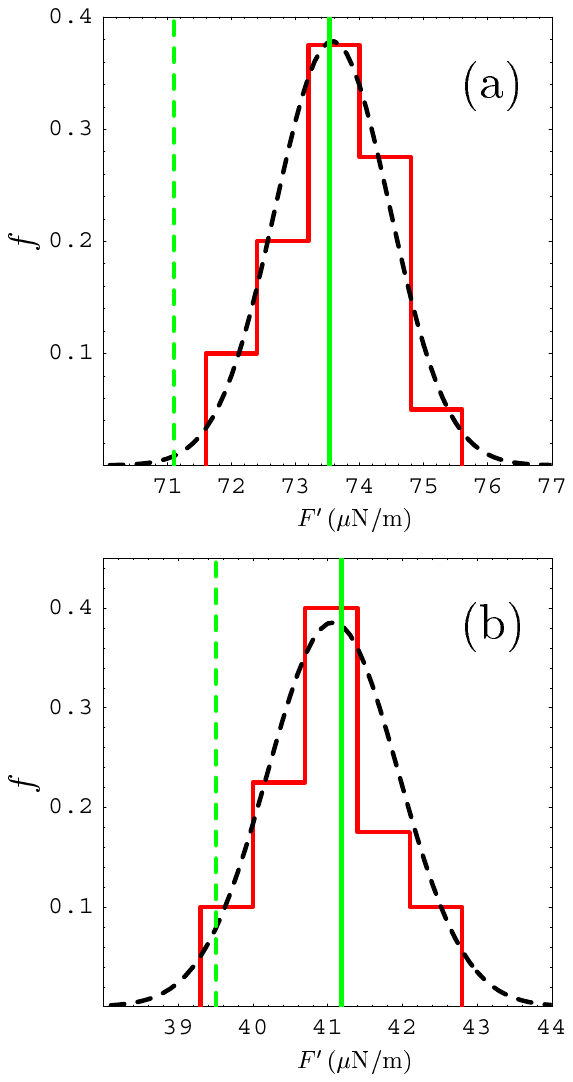}
}
\vspace*{-12.5cm}
\caption{(Color online) The histograms for the measured gradients of the
Casimir force at separations (a) $a=235\,$nm and (b) $a=275\,$nm.
$f$ is the fraction of 40 data points having the force values in the
bin indicated by the respective vertical lines of the histogram.
The corresponding Gaussian distributions are shown by the dashed black
lines. The solid and dashed vertical lines show the theoretical
predictions from the plasma and Drude model approaches, respectively.
}
\end{figure*}
\begin{figure*}[h]
\vspace*{-12.cm}
\centerline{\hspace*{1cm}
\includegraphics{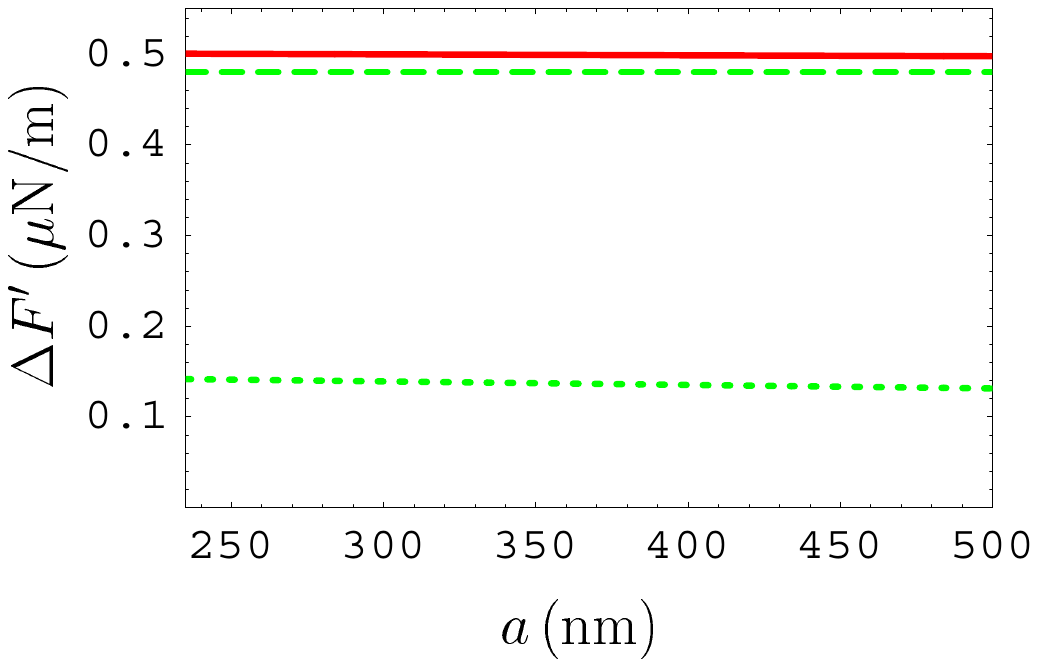}
}
\vspace*{-6.5cm}
\caption{(Color online) The random, systematic and total experimental
errors in the measured gradients of the Casimir force determined at
a 67\% confidence level are shown by the short-dashed, long-dashed and
solid lines, respectively. The measurement scheme with applied
compensating voltage is used.
}
\end{figure*}
\begin{figure*}[h]
\vspace*{-7.cm}
\centerline{\hspace*{1cm}
\includegraphics{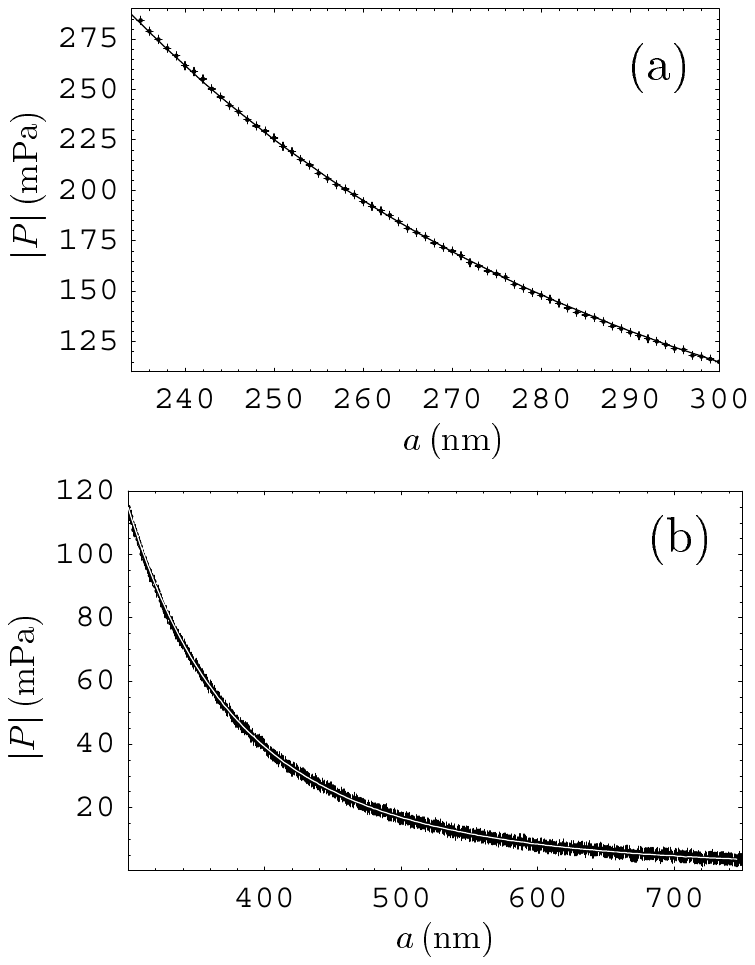}
}
\vspace*{-11.5cm}
\caption{Magnitudes of the mean Casimir pressure previously
measured\cite{19,20} are shown by (a) black and (b) white lines as
functions of separation. Magnitudes of the mean Casimir pressure
measured here are indicated as crosses. The arms of the crosses
are determined by errors in the measurement of separations and
force gradients.
}
\end{figure*}
\begin{figure*}[h]
\vspace*{-7.cm}
\centerline{\hspace*{1cm}
\includegraphics{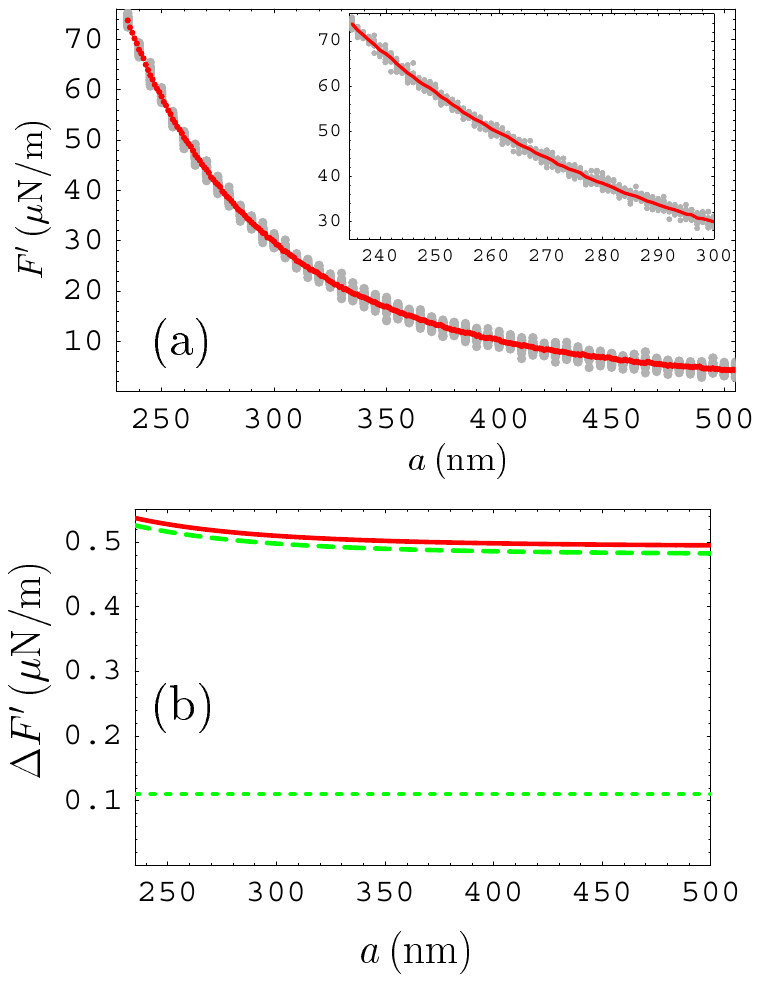}
}
\vspace*{-11.5cm}
\caption{(Color online)
(a)  Mean measured gradients of the Casimir force
as a function of separation are shown by solid lines.
Grey dots indicate all 40 individual force gradients plotted with the step
of 5\,nm (1\,nm in the inset).
(b)  The random, systematic and total experimental
errors in the measured gradient of the Casimir force determined at
a 67\% confidence level are shown by the short-dashed, long-dashed and
solid lines, respectively. The measurement scheme with different applied
voltages is used.
}
\end{figure*}
\begin{figure*}[h]
\vspace*{-8.cm}
\centerline{\hspace*{1cm}
\includegraphics{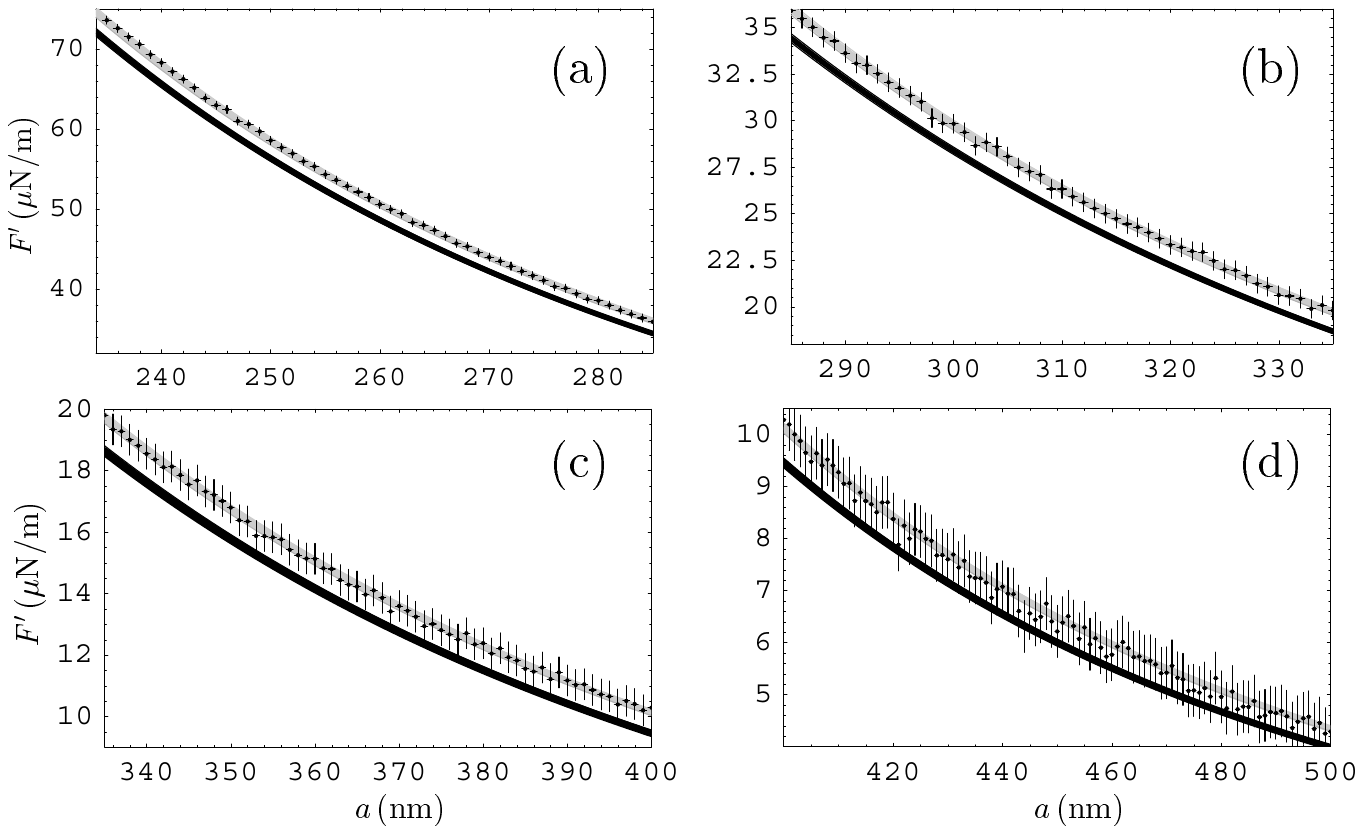}
}
\vspace*{-12.5cm}
\caption{Comparison between the experimental data for the
gradient of the Casimir force (crosses plotted at a 67\%
confidence level) and theory (black and grey bands computed using
the Drude and plasma model approaches, respectively) within
different separation regions.
The experimental data are obtained with the compensating voltage
applied to the plate.
}
\end{figure*}
\begin{figure*}[h]
\vspace*{-8.cm}
\centerline{\hspace*{1cm}
\includegraphics{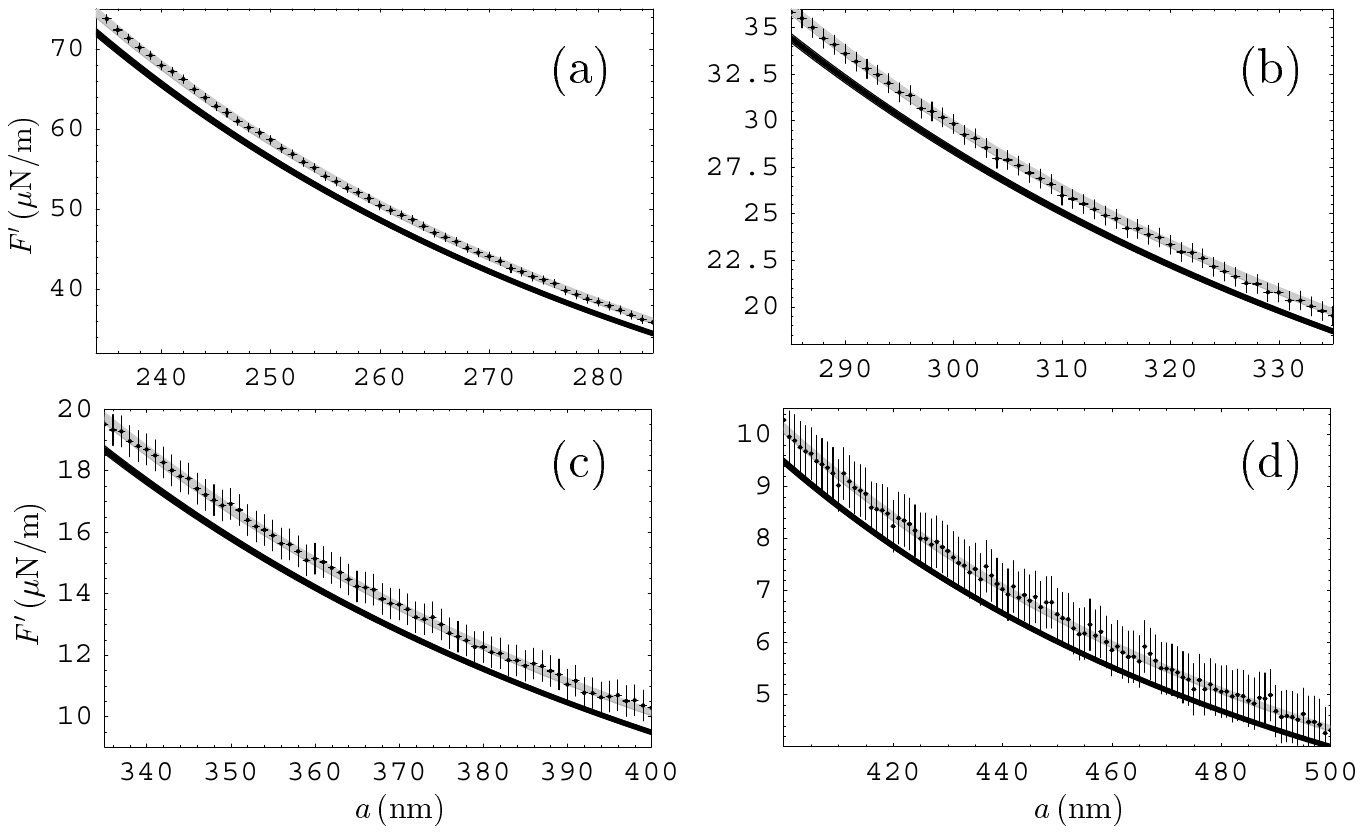}
}
\vspace*{-12.5cm}
\caption{Comparison between the experimental data for the
gradient of the Casimir force (crosses plotted at a 67\%
confidence level) and theory (black and grey bands computed using
the Drude and plasma model approaches, respectively) within
different separation regions.
The experimental data are obtained with different voltages
applied to the plate.
}
\end{figure*}
\begin{figure*}[h]
\vspace*{-12.cm}
\centerline{\hspace*{1cm}
\includegraphics{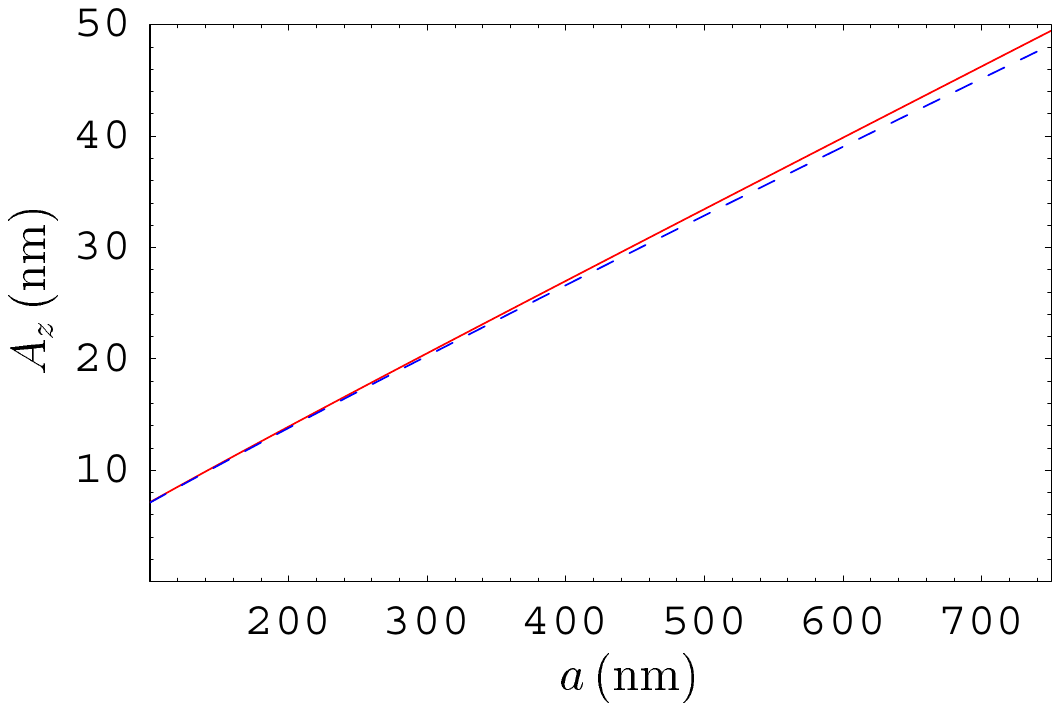}
}
\vspace*{-6.5cm}
\caption{Maximum allowed amplitudes of oscillations of the AFM
cantilever in the linear regime as a function of separation
are shown by the solid line (the plasma model approach) and
by the dashed line (the Drude model approach). The allowed
regions of $(a,A_z)$-plane lie beneath the lines.
}
\end{figure*}
\begingroup
\squeezetable
\begin{table}
\caption{The mean values of the gradients of the Casimir force
together with their total experimental errors
at different separations (column 1) measured in this work with
applied compensating voltage (column 2) and with different applied
voltages (column 3). Column~4 contains the mean gradients of the
Casimir force and their total experimental errors
obtained from the previously measured
pressures.\cite{19,20}
}
\begin{ruledtabular}
\begin{tabular}{cccc}
&\multicolumn{3}{c}{Gradients of the Casimir force
$F^{\prime}\,(\mu\mbox{N/m})$}
\\[1mm]
\cline{2-4}
$a$& measurements with & measurements with&
\\[-2.5mm]
(nm)&applied $V_0$ & different applied $V_i$ &IUPUI\\
\hline
236& $72.56\pm 0.50$ & $72.35\pm 0.54$ &$72.22\pm 0.34$ \\
240& $68.27\pm 0.50$ & $67.92\pm 0.53$ &$67.91 \pm 0.32$\\
250& $58.55\pm 0.50$ & $58.62\pm 0.53$ &$58.43 \pm 0.29$\\
260& $50.57\pm 0.50$ & $50.42\pm 0.52$ &$50.57 \pm 0.27$\\
270& $43.98\pm 0.50$ & $44.08\pm 0.52$ &$44.01\pm 0.25$ \\
280& $38.55\pm 0.50$ & $38.38\pm 0.51$ &$38.47 \pm 0.23$\\
290& $33.63\pm 0.50$ & $33.60\pm 0.51$ &$33.78 \pm 0.22$\\
300& $29.83\pm 0.50$ & $ 29.83\pm 0.51$ &$29.79 \pm 0.21$\\
350& $16.80\pm 0.50$ & $16.92\pm 0.50$ &$16.77\pm 0.18$ \\
400& $10.28\pm 0.50$ & $10.28\pm 0.50$ &$10.17 \pm 0.17$\\
450& $~6.22\pm 0.50$ & $~6.54\pm 0.50$ &$~6.53 \pm 0.16$\\
500& $~4.29\pm 0.50$ & $~4.32\pm 0.49$ &$~4.36 \pm 0.16$ \\
550& $~3.20\pm 0.50$ & $~2.87\pm 0.49$ &$~3.03 \pm 0.16$ \\
600& $~2.51\pm 0.50$ & $~2.12\pm 0.49$ &$~2.18 \pm 0.16$ \\
650& $~1.74\pm 0.50$ & $~1.56\pm 0.49$ &$~1.61 \pm 0.16$ \\
700& $~1.16\pm 0.50$ & $~1.17\pm 0.49$ &$~1.23 \pm 0.16$ \\
746& $~1.06\pm 0.50$ & $~0.82\pm 0.49$ &$~0.94 \pm 0.16$
\end{tabular}
\end{ruledtabular}
\end{table}
\endgroup
\end{document}